\documentclass[a4paper,11pt]{article}

\usepackage{enumitem}
\usepackage{float}

\usepackage{mathtools}

\usepackage{fullpage}
\usepackage[T1]{fontenc}

\usepackage{amssymb,amsmath,cite}

\usepackage{xcolor,bbold,comment}

\usepackage{amsthm}

\def\be{\begin{equation}\begin{aligned}}
\def\ee{\end{aligned}\end{equation}}
\def\nn{\nonumber}

\def\thefootnote{\arabic{footnote}}
\allowdisplaybreaks

\numberwithin{equation}{section}

\begin{document}

{\ }
\vspace{1.6cm}

\begin{center}

\def\thefootnote{\fnsymbol{footnote}}

{\Large \bf 
Scale-separated AdS$_3\times$S$^1$ vacua from IIA orientifolds 
}

\vskip 1.5cm

Fotis Farakos$^{1}$ and Matteo Morittu$^{2,3}$

\vskip 1cm
       {\it  $^{1}$Physics Division, National Technical University of Athens \\
        15780 Zografou Campus, Athens, Greece} 
        
        	\vspace{15pt}

        {\it  $^{2}$Departamento de F\'isica, Universidad de Oviedo, \\ 
        Avda. Federico Garc\'ia Lorca 18, 33007 Oviedo, Spain}

        \vspace{7pt}

        {\it  $^{3}$Instituto Universitario de Ciencias y Tecnolog\'ias Espaciales de Asturias (ICTEA), \\ 
        Calle de la Independencia 13, 33004, Oviedo, Spain}
\vspace{15pt}

\vspace{.5cm}

\vspace{1.3cm}

ABSTRACT 

\end{center}

We study supersymmetric AdS$_3$ flux vacua of massive type-IIA supergravity on anisotropic G2 orientifolds. Depending on the value of the $F_4$ flux the seven-dimensional compact space can either have six small and one large dimension such that the ``external'' space is scale-separated and effectively four-dimensional, or all seven compact dimensions small and parametrically scale-separated from the three external ones. Within this setup we also discuss the Distance Conjecture (including appropriate D4-branes), and highlight that such vacua provide a non-trivial example of the so-called Strong Spin-2 Conjecture.

\thispagestyle{empty} 
\setcounter{page}{0}

\baselineskip 6.0mm

\newpage

\tableofcontents

\section{Introduction}

Deducing the existence of an effectively lower-dimensional theory of gravity from ten-dimensional (or eleven-dimensional) string/M-theory is both pivotal for realistic model building and also presents an interesting theoretical question on its own right. 
In particular, supersymmetric AdS vacua with scale separation are interesting for a variety of reasons: for example, in some cases such vacua can be uplifted to de Sitter \cite{Kachru:2003aw}, while in principle they are also of interest in holography. 

The status of scale-separated AdS remains however unclear, 
even though one would expect supersymmetry to provide an advantage. 
For instance, the classical constructions with scale separation that appeared in \cite{DeWolfe:2005uu} (see also \cite{Behrndt:2004mj,Derendinger:2004jn,Lust:2004ig,Camara:2005dc,Caviezel:2008ik} for closely related work) invoke type-IIA supergravity with Romans mass and (smeared) O6-planes, 
the latter possibly being an essential ingredient of scale separation in any case \cite{Tsimpis:2012tu,Gautason:2015tig,Lust:2020npd}. 
The consistency of each of these ingredients or their combination has also been under debate \cite{Banks:2006hg,McOrist:2012yc,Lust:2019zwm}, while various steps in resolving or at least understanding some of the intricacies have been taken 
in \cite{Acharya:2006ne,Blaback:2010sj,Saracco:2012wc,Font:2019uva,Junghans:2020acz,Buratti:2020kda,Marchesano:2020qvg,Baines:2020dmu,DeLuca:2021mcj,Cribiori:2021djm,Andriot:2022yyj,Andriot:2022brg,Shiu:2022oti,Basile:2023rvm,Andriot:2023fss,Junghans:2023yue}. 
More recently, an analysis of the holographic duals was performed in \cite{Conlon:2021cjk,Apers:2022zjx,Apers:2022tfm,Quirant:2022fpn,Plauschinn:2022ztd,Apers:2022vfp}, and some extensions with anisotropies are discussed in \cite{Carrasco:2023hta,Tringas:2023vzn}. 
Controlled type-IIB vacua with scale separation are not known classically \cite{Petrini:2013ika,Emelin:2021gzx}, 
and the only known examples are the ones that require quantum corrections \cite{Kachru:2003aw} 
(and have a series of open issues \cite{Danielsson:2018ztv,Gautason:2018gln,Hamada:2018qef,Gao:2020xqh}), 
while there seems to be a scaling argument behind this difficulty \cite{Emelin:2020buq}. 
Arguments from supergravity also indicate the difficulty in constructing scale-separated AdS with extended supersymmetry \cite{Green:2007zzb,Cribiori:2022trc,Cribiori:2023gcy,Cribiori:2023ihv}\footnote{See also \cite{Apruzzi:2021nle,Apruzzi:2019ecr} for restrictions on scale separation in higher dimensional AdS spaces.}.  
In the mean time AdS$_3$ constructions with minimal supersymmetry and scale separation (but still with the same ingredients) 
have appeared in the literature \cite{Farakos:2020phe,Emelin:2022cac,VanHemelryck:2022ynr,Farakos:2023nms}\footnote{Recent work on type-II AdS$_3$ flux vacua with minimal or extended supersymmetry but without scale separation can be found, for example, in \cite{Dibitetto:2018ftj,Passias:2020ubv,Macpherson:2021lbr}.}. 
For a pedagogical and up-to-date review of the status of flux vacua with moduli stabilization and scale separation see e.g. \cite{VanRiet:2023pnx,Coudarchet:2023mfs}. 

In this work we will assume that smeared O6-planes are a consistent ingredient of flux compactifications and that there does exist an actual solution in string theory that is described in the low energy regime by such an approximation. 
In our constructions we will therefore build on the G2 orientifolds of \cite{Farakos:2020phe} which originally give rise to scale-separated AdS$_3$ vacua with a seven-dimensional (almost fully) isotropic compact internal space. 
However, our aim is to create a large anisotropy in the internal space such that six of the internal dimensions shrink (always maintaining large volume), while only one becomes large and comparable in magnitude to the typical length scale of the non-compact three-dimensional  (AdS$_3$) external space. 
In this way the actual external space is four-dimensional whereas the six-dimensional internal one is scale-separated (not parametrically though) from the external one. 
When instead the $F_4$ flux is made parametrically large, one recovers a seven-dimensional small internal compact space, albeit anisotropic, together with the AdS$_3$ external one.

\section{Massive type-IIA on G2 orientifolds}

We will work within the general setup discussed in \cite{Farakos:2020phe,Farakos:2023nms}; therefore, here, we only need to mention the salient features of such flux compactifications and establish our conventions. (Let us also note right away that we fix $\alpha'=1$). 

The setup is massive type-IIA supergravity compactified on a G2 space, which we take to be a toroidal orbifold $X_7 = T^7/(Z_2)^3$ with coordinates $\{y_i\}_{i=1,\dots,7}$.  
The $Z_2$ involutions generate a group $\Gamma=\{\Theta_\alpha,\Theta_\beta,\Theta_\gamma\}$, 
and are defined as 
\be
\begin{aligned}
\Theta_\alpha : (y^1, \dots, y^7 ) & \to (-y^1, -y^2, -y^3, -y^4, y^5, y^6, y^7) \, , 
\\
\Theta_\beta : (y^1, \dots, y^7 ) & \to (-y^1, -y^2, y^3, y^4, -y^5, -y^6, y^7) \, ,
\\
\Theta_\gamma : (y^1, \dots, y^7 ) & \to (-y^1, y^2, -y^3, y^4, -y^5, y^6, -y^7) \, ; 
\end{aligned}
\ee
one should also include in $\Gamma$ all their combinations, e.g. $\Theta_{\alpha\beta} = \Theta_{\alpha}\Theta_{\beta}$, etc.. 
Since the compact space is seven-dimensional, a three-dimensional non-compact space is left, which in \cite{Farakos:2020phe,Farakos:2023nms} is AdS$_3$. 
Eventually the ten-dimensional Einstein-frame metric involved in the compactification has the form 
\be
{\rm{d}}s^2_{10} = ({\text{Vol}}(X_7))^{-2} {\rm{d}}s^2_3 + \sum_{i=1}^7 r_i^2 {\rm{d}} y_i^2 \,, \quad i = 1,\dots,7 \,, 
\ee
and $\text{Vol}(X_7) = \prod_{i=1}^7 r^i$. 
Such G2 compactification would in principle preserve four real Killing spinors\footnote{More concretely, one can think of this in terms of gravitini such that the counting is valid also when supersymmetry is spontaneously broken.}, 
but, because one also includes mutually supersymmetric O2/O6-planes in these constructions, the number of supersymmetry is further reduced by half. 
The O2-planes fill the non-compact space, 
whereas the O6-planes have three dimensions that fill the non-compact space while the remaining four wrap 4-cycles 
in the seven-dimensional compact space as follows: 
\begin{align}
\begin{pmatrix} 
{\rm O}6_{\alpha}: & \times & \times & \times & \times & - & - & -  \\
{\rm O}6_{\beta}: & \times & \times & - & - & \times & \times & -  \\
{\rm O}6_{\gamma}: & \times & - & \times & - & \times & -& \times  \\
{\rm O}6_{\alpha\beta}: & - & - & \times & \times & \times & \times & -  \\ 
{\rm O}6_{\beta\gamma}: & - & \times & \times & - & - & \times & \times  \\ 
{\rm O}6_{\gamma\alpha}: & - & \times & - & \times & \times & - & \times  \\ 
{\rm O}6_{\alpha\beta\gamma}: & \times & - & - & \times & - & \times & \times 
\end{pmatrix} \, ,  
\end{align}
where the ``$\times$'' indicates a direction parallel to the given O6-plane and the ``$-$'' a direction orthogonal to it. 
The O6-planes would be localized in a full string theory solution, but here we consider them to be smeared along those directions. 
The full construction works nicely making the O6-planes actually the images of the O2-planes under the G2 involutions. 

Due to the various ingredients the only fluxes that can have non-zero background values are the $F_4$ flux, the $H_3$ flux and the Romans mass $F_0$, which are properly expanded on the basis of harmonic forms with the correct parities. The harmonic three-forms
\be
\Phi_i = \{ {\rm{d}}y^{127}, - {\rm{d}}y^{347}, - {\rm{d}}y^{567},  {\rm{d}}y^{136}, - {\rm{d}}y^{235},  {\rm{d}}y^{145},  {\rm{d}}y^{246} \} 
\ee 
provide a proper basis for the $H_3$ background flux and the harmonic four-forms $\Psi_i$, which are defined by 
\be 
\int_{X_7} \Phi_i \wedge \Psi_j =\delta_{ij} \, , 
\ee  
provide the proper basis for the $F_4$ flux.
In the mean time the only allowed moduli are the dilaton and the seven radii $r^i$ of the toroidal orbifold or, equivalently, the volumes of the seven 3-cycles $\Phi_i$, which we will indicate as $s^i$ and are defined as 
\be
s^1 = r^1r^2r^7 ,
s^2 = r^3r^4r^7 ,
s^3 = r^5r^6r^7 , 
s^4 = r^1r^3r^6 , 
s^5 = r^2r^3r^5 , 
s^6 = r^1r^4r^5 ,  
s^7 = r^2r^4r^6 . 
\ee
All the other closed string scalars are automatically truncated due to the orbifold/orientifold action. We ignore the open string moduli here: we assume that, being compact, they will be stabilized independently at their supersymmetric positions. 

In our analysis it is favorable to work with the moduli $x$, $y$ (not to be confused with the $y^i$ coordinates of the internal space) 
and $\tilde s^a$ (with $a=1,\dots,6$), which have the following properties: the moduli $x$ and $y$ are a combination of the overall compact space volume $\text{Vol}(X_7)$ and the dilaton $\phi$, namely
\be
x = - \frac{3 \sqrt{7}}{8} \phi + \frac{1}{2\sqrt{7}} \log[\text{Vol}(X_7)] \,, \quad 
y = - \frac{3}{2} \log[\text{Vol}(X_7)] - \frac{1}{8} \phi \,. 
\ee 
Then, since we treat the volume $\text{Vol}(X_7)$ as an independent modulus, one of the seven moduli $s^i$ is made redundant 
and can be eliminated by working with the six $\tilde s^a$ scalars that are deduced from the relations 
\be
\label{tilde-s} 
s^i = \text{Vol}(X_7)^{3/7} \tilde s^i \ ,  \ \ \text{with}\ i=1,\dots,7 
\,, \quad 
\tilde s^7 = \prod_{a=1}^6 \frac{1}{\tilde s^a} \ , \ \ \text{with}\ a=1,\dots,6 \,. 
\ee
To recap, we have six independent ``shape'' moduli $\tilde s^a$ (with $a=1,\dots,6$), 
while a seventh shape modulus, 
$\tilde s^7$, is given in terms of the other six. 

Having discussed the moduli, let us turn to the tadpole conditions. 
The contribution of the O2-planes to the tadpole will be cancelled by appropriately distributed D2-branes, 
whereas the O6-plane contribution will be cancelled by fluxes. 
Therefore, from the Bianchi identity $\rm{d}F_6=0$ we have 
\be
0 = \int_{X_7} H_3 \wedge F_4  \, 
\ee
due to vanishing net D2 charge. 
When integrating over each $i^{\rm th}$ 3-cycle, the Bianchi identity $\rm{d}F_2=0$ is given by 
\be
0= \int_i H_3 \wedge F_0 - 2 \pi \times 16  \, , 
\ee
where we have taken into account that $N_{\rm O6} = 2^3$ per 3-cycle. 
This setup has been analyzed extensively in \cite{Farakos:2020phe,Farakos:2023nms}: here, 
we refrain from getting into more details and refer the reader directly to the aforementioned articles. 

With these ingredients the direct dimensional reduction of the ten-dimensional Einstein-frame action down to three dimensions produces the bosonic sector 
\begin{equation}
\label{3D-bos}
e^{-1}\mathcal{L}_{3} = \tfrac{1}{2}R_3 -\tfrac{1}{4}(\partial x)^2  -\tfrac{1}{4}(\partial y)^2  
-\tfrac{1}{4} \sum_{i=1}^7 \frac{1}{(\tilde s^i)^2} \partial \tilde{s}^i\partial \tilde{s}^i 
- V(x,y,\tilde s) \,, 
\end{equation}
where $V(x,y,\tilde s)$ indicates the three-dimensional scalar potential. 
The latter can also be described by a superpotential that can be found in \cite{Farakos:2020phe} and will be presented momentarily. 
When deducing the metric of the moduli space of \eqref{3D-bos} one should not forget the restrictions \eqref{tilde-s}.

\section{AdS$_3 \times$S$^1$ scale-separated vacua}

\subsection{Supersymmetric conditions}

In this section, always referring to the base of harmonic forms, we work with the specific flux choices
\be
H_3 = h \sum_{i=1}^7 \Phi_i \,, \quad  F_4 = f \left( \Psi_1 + \Psi_2 + \Psi_3 - 3 \Psi_4 \right) + q \left( \Psi_5 + \Psi_6 - 2 \Psi_7 \right)  \,, 
\ee
such that $H_3 \wedge F_4 \equiv 0$ and the O2-tadpole cancels indeed with D2-branes. We also include a Romans mass $F_0=m$. 
Let us note that the parameters $f$ and $q$ can be made arbitrarily large and are not bounded by the tadpoles. Moreover, as far as our analysis is concerned, we are going to assume, consistently both with the tadpole cancellation condition and flux quantization, that $f,q>0$ and that $m,h<0$. 
This choice of signs for the fluxes is forced upon us, among other reasons, due to moduli stabilization. We could also have $f,q<0$ and $m,h>0$, which is the same setup, but we could not achieve moduli stabilization in our numerical examples if we assumed $m>0$ and $h<0$ (or vice-versa). 

In order to achieve scale separation in the large volume and weak coupling regime we are going to use the large $f$ limit, while we are going to take $q$ large to make the seventh radius, i.e. $r_7$, considerably bigger than the remaining compact dimensions. 
In that case four dimensions will be large and external (in the sense that the $r_7$ KK-modes can not be ignored for a consistent low-energy EFT) while the other six will be made considerably smaller with a clear separation of the corresponding scales. 

The flux quantization conditions, for $\alpha'=1$, are 
\be
h = (2 \pi)^2 K \,, \quad m = (2 \pi)^{-1} M \,, \quad f = (2 \pi)^3 N  \,, \quad q = (2 \pi)^3 Q \,, 
\ee
and from the cancellation of the O6-tadpole without any D6-branes we simply have $KM=16$. 
As we noted earlier, in our setup we have $m,h<0$, which means that $K,M<0$; we will therefore sometimes denote $-K=|K|$ and $-M=|M|$. 

For this setup the three-dimensional scalar potential can be written in terms of a real superpotential $P$ that was derived in \cite{Farakos:2020phe}. 
Then, in terms of the $\tilde s^a$ moduli (with $a=1,\dots,6$) the superpotential is 
\be\label{SuperP}
\begin{aligned}
\frac{P}{(2 \pi)^7} = & \  \frac{m}{8} \exp\left[\frac{y}{2} - \frac{\sqrt{7} x}{2}\right] 
+ \frac{h}{8}  \exp\left[ y + \frac{x}{\sqrt{7} }\right] \left( \sum_{a=1}^6 \frac{1}{\tilde s^a} + \prod_{a=1}^6 \tilde s^a \right) +
\\
& + \frac18 \exp\left[ y - \frac{x}{\sqrt{7} }\right] \left[ f \left( \tilde s^1 + \tilde s^2 + \tilde s^3 - 3 \tilde s^4 \right) 
+ q \left(\tilde s^5 + \tilde s^6 - 2 \prod_{a=1}^6 \frac{1}{\tilde s^a} \right) \right] \,. 
\end{aligned}
\ee
Having in mind supersymmetric critical configurations, let us observe that, since on a supersymmetric vacuum the vacuum energy is $\langle V \rangle=-4P^2$, the AdS$_3$ length scale is characterized by $P$ as $L_{\rm AdS} = \frac{1}{\sqrt{\vert V \vert}} \sim P^{-1}$ (for a constant three-dimensional Planck mass, which here, and from now on, we set to be unit).

A supersymmetric vacuum requires to have $P_x=P_y=P_a=0$. 
Once we perform the variations, 
we implement an ansatz for the various shape moduli $\tilde s^a$, 
each one associated with the volume of a corresponding 3-cycle, of the form 
\be\label{ModuliAnsatz}
\langle \tilde s^1 \rangle = \langle \tilde s^2 \rangle = \langle \tilde s^3 \rangle = \sigma 
\,, \quad \langle \tilde s^4 \rangle = \rho 
\,, \quad \langle \tilde s^5 \rangle = \langle \tilde s^6 \rangle = \tau \,. 
\ee
We further define 
\be
\mu = \frac{m}{f} \exp\left[ - \frac{y_0}{2} - \frac{5 x_0}{2 \sqrt{7}}\right]  \,, \quad  
\chi  = \frac{h}{f} \exp\left[ \frac{2 x_0}{\sqrt{7}}\right]  \,, \quad  
\gamma  = \frac{q}{f} \,, 
\ee
simply denoting $\langle x \rangle = x_0$ and $\langle y \rangle = y_0$. 
Let us furthermore remind the reader that  $\langle \tilde s^7 \rangle = 1 / \prod_{a=1}^6 \langle \tilde s^a \rangle$. 
Then, with the above definitions the supersymmetric equations lead to the following independent set of conditions: 
\be
\label{SUSY-eqs1}
\begin{aligned}
0&= - \frac{7 \mu}{2} - 3 \sigma - 2 \gamma \tau + \frac{3 \chi}{\sigma} + \frac{2 \chi}{\tau} 
+ \frac{\chi+\frac{2 \gamma}{\sigma^3 \tau^2}}{\rho}  + \rho \, (\sigma^3 \tau^2 \chi+3) \,, 
\\
0&= \frac{\mu}{2} + 3 \sigma + 2 \gamma \tau + \frac{3 \chi}{\sigma} + \frac{2 \chi}{\tau} 
+ \frac{\chi-\frac{2 \gamma}{\sigma^3 \tau^2}}{\rho}  + \rho \, (\sigma^3 \tau^2 \chi-3) \,, 
\end{aligned}
\ee
and 
\be
\label{SUSY-eqs2}
\begin{aligned}
0&= 1 + \frac{2 \gamma}{\rho \sigma^4 \tau^2} - \frac{\chi}{\sigma^2} + \rho \sigma^2 \tau^2 \chi  \,, 
\\
0&= -3 + \frac{2 \gamma}{\rho^2 \sigma^3 \tau^2} - \frac{\chi}{\rho^2} + \sigma^3 \tau^2 \chi  \,, 
\\
0&= \gamma + \frac{2 \gamma}{\rho \sigma^3 \tau^3} - \frac{\chi}{\tau^2} + \rho \sigma^3 \tau \chi \,. 
\end{aligned}
\ee

Once the values of the various parameters have been determined, we can extract 
\be
x_0 = \frac{\sqrt{7}}{2} \log\left[\frac{f \chi}{h}\right] 
\,, \quad 
y_0 = 2 \log\left[\frac{m}{f \mu}\right] - \frac{5}{2} \log\left[ \frac{f \chi}{h} \right] \,, 
\ee
which in turn give the dilaton $\phi$ and the volume $\text{Vol}(X_7)$. In particular,
\be
\label{GS-VOL}
g_s = e^{\langle \phi \rangle} =  \frac{h }{f^{3/4} \chi }  \left(\frac{\mu}{m}\right)^{1/4} 
\,, \quad 
\text{Vol}(X_7) = f^{49/16} \left(\frac{\chi}{h}\right)^{7/4} \left(\frac{\mu}{m} \right)^{21/16} \,. 
\ee
It is clear from the above expressions that, when $f$ is large enough, the volume can be made large and the coupling can be made weak, even though one has to check the radii one-by-one and make sure that each one is independently large, in case the space is very anisotropic. 
We will do this in a while; for the moment, let us simply note that the actual volumes of the 3-cycles are given by $s^i = \text{Vol}(X_7)^{3/7} \tilde  s^i$. 

We can solve the two equations of \eqref{SUSY-eqs1}, which actually correspond to $P_x=0=P_y$, to deduce 
\be
\label{mu-chi}
\begin{aligned}
\mu & = \frac{3 (\rho - \sigma)}{2} 
+ \frac{\gamma}{\rho \sigma^3 \tau^2} 
- \gamma \tau \,, 
\\
\chi & = \frac{9 \rho (\rho - \sigma) \sigma^3 \tau^2 + 6 \gamma (1 - \rho \sigma^3 \tau^3)}{4 \sigma^2 \tau (2 \rho \sigma + 3 \rho \tau + \sigma \tau + \rho^2 \sigma^4 \tau^3)} \,. 
\end{aligned}
\ee
Substituting then \eqref{mu-chi} into \eqref{SUSY-eqs2} will give us three equations with four unknowns, i.e. ($\sigma,\rho,\tau,\gamma$): as a result, we have one free parameter which can be conveniently chosen to be $\gamma$ as it controls the relative anisotropy within the $F_4$ flux components. 
It is precisely this freedom that will allow us to break scale separation in one compact dimension and thus make the external/large space effectively four-dimensional. 
The three equations that have to be solved are then 
\be
\label{EqsRTS}
\begin{aligned}
0&= 1 
+ \frac{2 \gamma}{\rho \sigma^4 \tau^2} 
+ \frac{
3 (\rho \sigma^4 \tau^2 -1) [3 \rho \sigma^3 \tau^2 (\rho - \sigma) + 2 \gamma (1- \rho \sigma^3 \tau^3) ]
}{
4 \sigma^4 \tau (2 \rho \sigma + 3 \rho \tau + \sigma \tau + \rho^2 \sigma^4 \tau^3) 
} \,, 
\\
0&= -3 
+ \frac{2 \gamma}{\rho^2 \sigma^3 \tau^2} 
+ \frac{3
\left(\sigma^3 \tau^2 - \frac{1}{\rho^2}\right) 
[3 \rho \sigma^3 \tau^2 (\rho - \sigma) + 2 \gamma (1- \rho \sigma^3 \tau^3) ]
}{
4 \sigma^2 \tau (2 \rho \sigma + 3 \rho \tau + \sigma \tau + \rho^2 \sigma^4 \tau^3) 
} \,, 
\\
0&= \gamma 
+ \frac{2 \gamma}{\rho \sigma^3 \tau^3} 
+ \frac{3
\left(\rho \sigma^3 \tau^3 - 1 \right) 
[3 \rho \sigma^3 \tau^2 (\rho - \sigma) + 2 \gamma (1- \rho \sigma^3 \tau^3) ]
}{
4 \sigma^2 \tau^3 (2 \rho \sigma + 3 \rho \tau + \sigma \tau + \rho^2 \sigma^4 \tau^3) 
} \,. 
\end{aligned}
\ee
Ideally we would like to solve the equations in \eqref{EqsRTS} analytically and express the solutions as functions of $\gamma$, that is $\sigma=\sigma(\gamma)$, $\rho=\rho(\gamma)$ and $\tau=\tau(\gamma)$, and further deduce from \eqref{mu-chi} $\mu=\mu(\gamma)$ and $\chi=\chi(\gamma)$. However, due to the highly nonlinear polynomial structure of \eqref{EqsRTS} it does not seem possible to directly find its analytic solutions; therefore, from now on we will turn to numerical methods.

\subsection{Proceeding numerically} 
\label{Numerics}

As we already wrote, our aim is to achieve a large volume and weak coupling regime where the seventh radius $r_7$ is made large, namely of the same order of magnitude of the AdS$_3$ radius (or even bigger), and the remaining six radii $r_a$ remain small.  
The comparison between the relevant KK-mode scales associated with each of the radii $r_i$ and the AdS length scale $L_{\rm AdS}$ is made by a formula that is derived, for example, in \cite{Farakos:2023nms}: for the lowest-lying KK-mode of a tower of states related to some radius $r_i$, 
\be
\label{KKmass}
m_{{\rm KK},i}^2 = \frac{(2 \pi)^2 }{(\text{Vol}(X_7))^2 \ r_i^2} \times \left( \frac{(2 \pi)^7}{2} \right)^2 \, 
\ee
and, as a consequence,
\be
\frac{L_{\rm{KK},i}^2}{L_{\rm AdS}^2} = \frac{|\langle V \rangle|}{m_{{\rm KK},i}^2} 
= \frac{4}{\pi^2} \, \left(\frac{P}{(2 \pi)^7}\right)^2 \, (\text{Vol}(X_7))^2 \, r_i^2 \,, 
\ee
where $|\langle V \rangle|=1/L_{\rm AdS}^2$.
As we are going to see exploring what happens for different values of $\gamma$ and $f$, it is indeed possible to achieve
\be
\frac{L_{\rm{KK},a}^2}{L_{\rm AdS}^2} \rightarrow \, 0 
\,, \quad 
\frac{L_{\rm{KK},7}^2}{L_{\rm AdS}^2} \gtrsim 1 
\,, \quad g_s \ll 1 
\,, \quad r_i \gg 1 
\,, 
\ee
so that the large external space becomes four-dimensional (again in the sense that the $r_7$ KK-modes cannot be ignored for a consistent EFT) and the small hidden space six-dimensional. 

We would like to emphasize that the anisotropy between the radius $r_7$ and the other six radii $r_a$ in our constructions is not parametric. 
This happens because such anisotropy is controlled by the parameter $\gamma$ that is consistent with moduli stabilization only for a finite range of values. Nevertheless, as we sill see, within those values we could find quite a large anisotropy. 
It is not clear of course whether there could be another choice of fluxes that can make the anisotropy which singles out $r_7$ parametric. (Such a setup does exist for the case of parametric scale separation between $r_{2,4,6}$ and the rest of the radii; it was studied in \cite{Farakos:2023nms}). 

We would also like to stress that, once $r_7$ is made of the same order of magnitude of $L_{\rm AdS}$, the large external space is not AdS$_4$, but AdS$_3\times$S$^1$, also underlying the fact that we have a supersymmetric vacuum with two (and not four) Killing spinors.

\subsubsection{Having $\gamma \sim {\cal O}(1)$ and (almost) no anisotropy} 

For values of $\gamma \sim \mathcal{O}$(1) it is easy to check that (almost) no anisotropy is generated between the seven radii $r_i$: the system therefore behaves essentially like in \cite{Farakos:2020phe}.
Explicitly, solving the equations in \eqref{EqsRTS} for $\gamma = 1$, we find 
\be
\sigma \approx 1.95 \,, \quad \rho \approx 0.18 \,, \quad \tau \approx 1.95 \,. 
\ee
Then, from \eqref{GS-VOL} with the proper numerical values of $\mu$ and $\chi$ evaluated from \eqref{mu-chi} we can extract the 
behaviour of the string coupling 
\be
g_s \approx 2.897 \times \frac{|K|}{N^{3/4} |M|^{1/4}} \,, 
\ee
and the behaviour of the radii 
\be
\frac{r_{1,3}}{\frac{2\pi N^{7/16}}{|M|^{3/16} |K|^{1/4}}} \approx 0.93 \,, \, \frac{r_{2,4}}{\frac{2\pi N^{7/16}}{|M|^{3/16} |K|^{1/4}}} \approx 0.95 \,, \, \frac{r_{5}}{\frac{2\pi N^{7/16}}{|M|^{3/16} |K|^{1/4}}} \approx 3.04 \,, \, \frac{r_{6}}{\frac{2\pi N^{7/16}}{|M|^{3/16} |K|^{1/4}}} \approx 0.29 \,,
\ee
and 
\be
\frac{r_{7}}{\frac{2\pi N^{7/16}}{|M|^{3/16} |K|^{1/4}}} \approx 3.04 \,.
\ee  
By appropriately choosing the flux units we can end up having large internal volume ${\rm Vol}(X_7)$ 
(it is sufficient to take $N^{7/16} |M|^{-3/16} |K|^{-1/4} \gg 1$) 
and a small coupling $g_s$. 
We also clearly notice that all the seven radii $r_i$ are comparable in magnitude and therefore (almost) no anisotropy is generated yet. 

In any case, since 
\be
\frac{L_{\rm{KK},i}^2}{L_{\rm AdS}^2} \approx 0.21 \times \frac{K^2 |M|}{N} \,, 
\ee
for parametrically large $N$ we can get parametric scale separation between the AdS$_3$ external space and the seven-dimensional internal one.
For example, consistently with the tadpoles, one can have $K = -16$, $M = - 1$ and $N = 10^4$, or any other parametrically large value of $N$. 
The fact that for parametrically large $N$ we recover AdS$_3$ as external space is a general feature of the construction we are presenting here.

\subsubsection{Intermediate $\gamma$ values}
\label{Intermediate}

To try to detach the seventh radius $r_7$ from the other six internal radii $r_a$ and make the former comparable in magnitude to the AdS length scale $L_{\rm AdS}$, we will now increase $\gamma$ and consider the cases $\gamma \sim \mathcal{O}(10^3)$ and $\gamma \sim \mathcal{O}(10^6)$. 
This choice opens a window of values of $N$ within which the above scenario can be realized. (When $\gamma$ becomes smaller than ${\cal O}(1)$, then $r_5$ is detached; we analyze this in the appendix).

For the aforementioned values of $\gamma$ the equations in \eqref{EqsRTS}, which determine the properties of the solution and verify moduli stabilization, are solved for the values of $\sigma$, $\rho$ and $\tau$ that we present in Table \ref{TAB1}. 
For such values, from \eqref{GS-VOL} and with the proper $\mu$ and $\chi$ evaluated thanks to \eqref{mu-chi}, we can consequently estimate $g_s$ to be
\be
\label{GS36} 
g_s \approx 1.92 \times \frac{|K|}{N^{3/4} |M|^{1/4}} 
\ee
(which is roughly the same for both the values of $\gamma$ under consideration). 
In addition, the radii take the values that we gather in Table \ref{TAB2}. 
From \eqref{GS36} and Table \ref{TAB2} we readily see that the large volume (i.e. large radii) and weak coupling regime can be realized by adopting appropriate choices of the flux units, e.g. by choosing (always in accordance with the tadpole cancellation condition) $K=-16$, $M=-1$ 
and $N$ sufficiently large.

\begin{table}
\begin{center}
\renewcommand{\arraystretch}{1.5}
\begin{tabular}{|c||c||c||c|}
     \hline 
     $\gamma$ & $\sigma$ & $\rho$ & $\tau$  \\ 
     \hline \hline
     $10^3$   &  $93.12$  & $0.02$ & $0.06$ \\
     \hline
     $10^6$  &  $4823.5$ & $0.0012$ & $0.0032$ \\
     \hline
\end{tabular}
\caption{
\label{TAB1} 
This table presents the critical values of $\sigma$, $\rho$ and $\tau$ as $\gamma$ takes the values $\gamma = 10^3$, $10^6$. 
}
\end{center}
\end{table}

\begin{table}
\begin{center}
\renewcommand{\arraystretch}{1.5}
\begin{tabular}{|c||c||c||c||c||c|}
     \hline 
     $\gamma$ & $\frac{r_{1,3}}{\frac{2\pi N^{7/16}}{|M|^{3/16} |K|^{1/4}}}$ & $\frac{r_{2,4}}{\frac{2\pi N^{7/16}}{|M|^{3/16} |K|^{1/4}}}$ & $\frac{r_{5}}{\frac{2\pi N^{7/16}}{|M|^{3/16} |K|^{1/4}}}$ & $\frac{r_{6}}{\frac{2\pi N^{7/16}}{|M|^{3/16} |K|^{1/4}}}$ & $\frac{r_{7}}{\frac{2\pi N^{7/16}}{|M|^{3/16} |K|^{1/4}}}$  \\ 
     \hline \hline
     $10^3$   &  $1.09$  & $0.94$ & $1.88$ & $0.54$ & $2821$ \\
     \hline
     $10^{6}$  & $1.086$  & $0.941$ & $1.881$ & $0.543$ & $2.822\times 10^6$ \\
     \hline
\end{tabular}
\caption{\label{TAB2} This table shows the values of the radii $r_i$ as $\gamma$ takes the values $\gamma = 10^3$, $10^6$. One can clearly notice that, as $\gamma$ becomes larger, $r_7$ becomes bigger than the other six radii $r_a$ (for fixed flux units).}
\end{center}
\end{table}

As already emphasized, the arbitrarily parametrically large $N$ limit always leads to full scale separation. At this stage, however, we would like to understand if for some moderate (but still sufficiently large) value of $N$ only six of the compact dimensions can remain small while the seventh one can become of the same order of magnitude of the AdS$_3$ length scale (or even larger). 
This means that we want to investigate whether
\be
\frac{L_{\rm{KK},7}^2}{L_{\rm AdS}^2} = \frac{4}{\pi^2} \left(\frac{P}{(2\pi)^7} \right)^2 (\text{Vol}(X_7))^2 \, r_7^2 \gtrsim  1
\,, \quad 
\frac{L_{\rm{KK},a}^2}{L_{\rm AdS}^2} \ll 1 
\,. 
\ee 
More precisely, we have the ratios 
\be
\frac{L_{\rm{KK},7}^2}{L_{\rm AdS}^2} \Big{|}_{\gamma = 10^3} \approx 6.418 \times 10^4 \times \frac{K^2 |M|}{N} 
\,, \quad 
\frac{L_{\rm{KK},7}^2}{L_{\rm AdS}^2} \Big{|}_{\gamma = 10^6} \approx 6.412 \times 10^{10} \times \frac{K^2 |M|}{N} \,, 
\ee
which offer a wide range of values of $N$ that can be checked. 
Indeed, when (once more) fixing $K = -16$ and $M = -1$, we find, for $\gamma = 10^3$, the results that are presented in Table \ref{TAB4}, and, for $\gamma = 10^6$, we find the values that are reported in Table \ref{TAB5}, together with the relative behaviour of the string coupling and the other radii. 
We observe that, as $\gamma$ increases, the possibility to decouple the seventh radius $r_7$ from the other six radii $r_a$ becomes more and more achievable, with a simultaneous extension of the range of the values of $N$ (once $K$ and $M$ have been fixed) we can refer to. 

\begin{table}
\begin{center}
\renewcommand{\arraystretch}{1.5}
\begin{tabular}{|c||c||c||c||c||c||c|}
     \hline 
     $$ &  $N$ & $g_s$ & $r_a$ & $r_7$ & $\frac{L_{\rm{KK},7}^2}{L_{\rm AdS}^2}$ & $\frac{L_{\rm{KK},a}^2}{L_{\rm AdS}^2}$ \\ 
     \hline \hline
     $(a)$  & $ 10^{10}$ & $9.72 \times 10^{-7} $ & ${\cal O}(10^4)$& $2.10 \times 10^8$  &  $1.64\times10^{-3}$ & ${\cal O}(10^{-10})$  \\
     \hline
     $(b)$  & $10^7$ & $1.73 \times 10^{-4}$ & ${\cal O}(10^3)$& $1.02\times10^7$ &  $1.64$ & ${\cal O}(10^{-7})$  \\
     \hline
     $(c)$  & $10^4$ & $0.03$ & $ {\cal O}(10) $& $4.98\times10^4$ & $1.64\times10^3$ & ${\cal O}(10^{-4})$  \\
     \hline
\end{tabular}
\caption{\label{TAB4} This table shows the three interesting regimes one can end up with while changing $N$ for $\gamma = 10^3$, once the other flux units have been fixed, namely $K=-16$ and $M=-1$. 
When a circumstance like (a) realizes, one has full scale separation; 
if, instead, one works with cases similar to (b) or (c), then the radius $r_7$ disentangles from the other six radii and the external space becomes effectively AdS$_3\times$S$^1$. 
For the in-between values of $N$ one gets of course intermediate results. 
Note that $N$ can not be too small in order for the large volume/weak coupling condition to still be satisfied.} 
\end{center}
\end{table}

\begin{table}
\begin{center}
\renewcommand{\arraystretch}{1.5}
\begin{tabular}{|c||c||c||c||c||c||c|}
     \hline 
     $$  &  $N$ & $g_s$ & $r_a$ & $r_7$ & $\frac{L_{\rm{KK},7}^2}{L_{\rm AdS}^2}$ & $\frac{L_{\rm{KK},a}^2}{L_{\rm AdS}^2}$ \\ 
     \hline \hline
     $(a)$  & $10^{19}$ & $1.73\times10^{-13}$ & ${\cal O}(10^8)$& $1.82\times10^{15}$ &  $1.64\times10^{-6}$ & ${\cal O}(10^{-19})$  \\
     \hline
     $(b)$  & $10^{13}$ &$5.46\times10^{-9}$ & $ {\cal O}(10^6) $& $4.32\times10^{12}$ & $1.64$ & ${\cal O}(10^{-13})$  \\ 
     \hline
     $(c)$  & $10^7$ &$1.73 \times 10^{-4}$ & $ {\cal O}(10^3) $& $1.02\times10^{10}$ & $1.64\times10^6$ & ${\cal O}(10^{-7})$  \\
     \hline
\end{tabular}
\caption{\label{TAB5} This table shows three interesting regimes one can end up with while changing $N$ for $\gamma = 10^6$, 
once the other flux units have been fixed, namely $K=-16$ and $M=-1$. 
One can again see a similar behaviour with respect to the case $\gamma=10^3$, which is presented in Table \ref{TAB4}: a pattern indeed emerges. 
For the in-between values of $N$ one gets of course intermediate results, while one can appreciate that the windows for the various regimes cover now a wider range of values of $N$.} 
\end{center}
\end{table}

\begin{table}
\begin{center}
\renewcommand{\arraystretch}{1.5}
\begin{tabular}{|c||c||c||c|}
     \hline 
     $\gamma$ & $\sigma$ & $\rho$ & $\tau$  \\ 
     \hline \hline
     $10^{9}$ & $249833$ & $0.0000555$ & $0.0001666$ \\
     \hline
     $2.5305 \times 10^9$ & $424673$ & $0.0000373$ & $0.0001119$\\
     \hline
\end{tabular}
\caption{\label{TAB6} This table exhibits the critical values of $\sigma$, $\rho$ and $\tau$ as $\gamma$ takes the values $\gamma=10^9$, $2.5305 \times 10^9$.} 
\end{center}
\end{table}

\begin{table}
\begin{center}
\renewcommand{\arraystretch}{1.5}
\begin{tabular}{|c||c||c||c||c||c|}
     \hline 
     $\gamma$ & $\frac{r_{1,3}}{\frac{2\pi N^{7/16}}{|M|^{3/16} |K|^{1/4}}}$ & $\frac{r_{2,4}}{\frac{2\pi N^{7/16}}{|M|^{3/16} |K|^{1/4}}}$ & $\frac{r_{5}}{\frac{2\pi N^{7/16}}{|M|^{3/16} |K|^{1/4}}}$ & $\frac{r_{6}}{\frac{2\pi N^{7/16}}{|M|^{3/16} |K|^{1/4}}}$ & $\frac{r_{7}}{\frac{2\pi N^{7/16}}{|M|^{3/16} |K|^{1/4}}}$  \\ 
     \hline \hline
     $10^{9}$ & $1.086$  & $0.941$ & $1.881$ & $0.543$ & $2.822\times 10^9$ \\
     \hline
      $2.5305 \times 10^9$ & $1.086$ & $0.941$ & $1.881$ & $0.543$ & $7.140\times 10^9$ \\
     \hline
\end{tabular}
\caption{\label{TAB7} This table shows the values of the radii $r_i$ as $\gamma$ takes the values $\gamma=10^9$, $2.3502 \times 10^9$. One can clearly notice that $r_7$ is much bigger than the other six radii $r_a$ for fixed flux units.}
\end{center}
\end{table}

\subsubsection{Large $\gamma$ values and maximal anisotropy} 

In this subsection we will work with $\gamma \sim {\cal O}(10^9)$. 
This is roughly the largest value that $\gamma$ can have while keeping moduli stabilization intact. 
Indeed, from our numerics we see that $\gamma \approx 2.5305 \times 10^9$ appears to (approximately) be the limiting value that $\gamma$ 
can take before solutions to the extremization of \eqref{SuperP} with the ansatz \eqref{ModuliAnsatz} cease to exist. 

Solving the equations in \eqref{EqsRTS} we get the results presented in Table \ref{TAB6}, 
together with
\be
g_s \approx 1.92 \times \frac{|K|}{N^{3/4} |M|^{1/4}} \,, 
\ee
and the values presented in Table \ref{TAB7} as far as the radii $r_i$ are concerned. 
We can also check that a consistent anisotropy, where only six of the compact dimensions remain small while the seventh one becomes of the same order of magnitude of the AdS$_3$ length scale (or even larger) can be achieved. 
Focusing, as a matter of illustration, on $\gamma = 10^9$, we search for a realization of
\be
\frac{L_{\rm{KK},7}^2}{L_{\rm AdS}^2} \approx 6.412 \times 10^{16} \times \frac{K^2 |M|}{N} \gtrsim  1 
\,, \quad 
\frac{L_{\rm{KK},a}^2}{L_{\rm AdS}^2} \approx 10^{-3} \times \frac{K^2 |M|}{N} \ll 1  
\,. 
\ee
We also choose to keep the same $K$ and $M$ units of flux as before, i.e. 
\be
K = -16 \,, \quad M = -1 \,,
\ee
so that the tadpole cancellation condition is satisfied and we can simply vary the value of $N$. 

In particular, if we take $N \sim 10^{19}$, we find (always in appropriate string-length units, since $\alpha'=1$) 
\be
g_s \sim 10^{-13} \,, \quad r_a \sim 10^8 \,, \quad r_7 \sim 10^{18} \,, 
\ee
and
\be
\frac{L_{\rm{KK},7}^2}{L_{\rm AdS}^2} \sim 1 
\,, \quad 
\frac{L_{\rm{KK},a}^2}{L_{\rm AdS}^2} \sim  10^{-19} \,. 
\ee
As we expect, because the characteristic length scale associated with $r_7$ is comparable to the AdS$_3$ length scale, the previous relation shows that the external space becomes effectively four-dimensional (that is, the $r_7$ KK-modes can not be ignored), while the internal space is six-dimensional and it is scale-separated from the large external one. 

It is also possible to choose values of $N$ such that a different separation of scales is achieved, of the form 
\be 
L_{\rm{KK},7}^2 \gg L_{\rm AdS}^2 \gg L_{\rm{KK},a}^2 \,. 
\ee
For example, when $N \sim 10^{9}$, 
\be
g_s \sim 10^{-6} \,, \quad r_a \sim 10^{4} \,, \quad r_7 \sim 10^{13}
\ee
and
\be
\frac{L_{\rm{KK},7}^2}{L_{\rm AdS}^2} \sim 10^{10} 
\ , \quad 
\frac{L_{\rm{KK},a}^2}{L_{\rm AdS}^2} \sim  10^{-9} \,. 
\ee 
The four-dimensional external space is again AdS$_3 \times$S$^1$, 
still well scale-separated from the six-dimensional internal one; however, since the radius of the seventh compact dimension is larger than the AdS length scale, one could say that the external space effectively becomes AdS$_3 \times \mathbb{R}^1$. 

We note once more that, when the integer $N$ takes parametrically large values, even though the compact space may remain highly anisotropic, scale separation makes the internal space seven-dimensional and the external space AdS$_3$. 

Moreover, let us observe that our flux choices do not exhaust all the possible options that would give such an anisotropy; we leave a general scan of the various possibilities for future work.

\section{Mass spectrum}  
\label{MassSpectrum}

We now turn to the study of the mass spectrum of the model of interest focusing on the closed string moduli that we considered until now. 
Once we divide the masses of the scalar fields $x$, $y$ and $\tilde s^a$ by the vacuum energy, we see that the analysis does not depend on the specific choice of the units $N$ of the $F_4$ flux, but only on $\mu$, $\chi$ and $\gamma$. 
Therefore here we will collect some results and considerations involving the mass spectrum of the model simply as $\gamma$ takes the values $\gamma = 1$, $10^3$, $10^6$ and $\gamma \sim \mathcal{O}(10^9)$. 
We should remember of course that the relative magnitude of the $r_7$ KK-mode masses compared to the AdS length scale depends on $N$. 
This means that depending on the value of $N$ the three-dimensional theory with only the $x$, $y$ and $\tilde s^a$ is either a consistent truncation of a four-dimensional theory or an actual three-dimensional effective theory. 

As anticipated, for the evaluation of the masses we will proceed by considering the AdS$_3$ construction with the eight moduli $x$, $y$ and $\tilde s^a$. We will firstly focus on the $\gamma=10^9$ case since it is the circumstance with the larger window for the values of $N$ and thus covers all the interesting cases.

The scalar potential takes the form 
\be
\frac{V}{(2 \pi)^{14}} = F(\tilde s^a) \, e^{2 y - \frac{2 x}{\sqrt 7}} 
+ H(\tilde s^a) \, e^{2 y + \frac{2 x}{\sqrt 7}} 
+ C \, e^{y - \sqrt 7 x} 
+ T(\tilde s^a) \, e^{\frac{3y}{2} - \frac{5 x}{2 \sqrt 7}} \, 
\ee
with
\be
\begin{aligned}
F & = \frac{1}{16} \left[ f^2 \left( \sum_{a=1}^3 (\tilde s^a)^2 + 9((\tilde s^4)^2) \right)
+ q^2 \left( \sum_{a=5}^6(\tilde s^a)^2 \right)
+ 4 q^2 \prod_{a=1}^6 (\tilde s^a)^{-2} \right]   \,, 
\\
H & = \frac{h^2}{16} \left( \sum_{a=1}^6 (\tilde s^a)^{-2} + \prod_{a=1}^6 (\tilde s^a)^{2}  \right)   \, 
\end{aligned} 
\ee
and 
\be
C = \frac{m^2}{16}  
\,, \quad
T  = - \frac{h m}8 \left( \sum_{a=1}^6 \frac{1}{\tilde s^a} + \prod_{a=1}^6 \tilde s^a \right)  \,. 
\ee
From here we evaluate the Hessian matrix and its eigenvalues: 
\be
\begin{aligned}
 \text{Eigen}\left[ \frac{\langle V_{IJ} \rangle}{\vert\langle V \rangle\vert} \right]  = & \{ 8.872 \times 10^8, 3.204 \times 10^7, 2.604 \times 10^7, -8.038, \\& 4.152, 1.720 \times 10^{-10}, 6.409 \times 10^{-11}, 6.399 \times 10^{-11} \} \,,
\end{aligned}
\ee
where the eight indices $I, J$ run over the $x$, $y$ and $\tilde s^a$. 
Moreover, once the fields have been properly normalized, we get
\be
m^2 L_{\rm AdS}^2  =  \text{Eigen}\left[ \langle K_{IJ} \rangle^{-1} \frac{\langle V_{IJ} \rangle}{\vert \langle V \rangle \vert} \right] 
\approx \{ 49.192, 8, 8, 8, 3.934, 2.095, 1.778, -0.998 \} \,,
\ee
the matrix $K_{IJ}$ being 
\be
K_{IJ} = 2 \times  \begin{bmatrix}
    1/4 & 0 & 0 \\
    0 & 1/4 & 0 \\
    0 & 0 & G_{ab}
  \end{bmatrix} \,,
\quad \text{where} 
\quad G_{ab} = \frac{1+\delta_{ab}}{4 \, \tilde s^a \tilde s^b} \,. 
\ee 
Let us importantly observe that, once the modes have been canonically normalized, the tachyon respects the Breitenlohner--Freedman (BF) bound, which for AdS$_3$ is 
\be
m^2 \geq \langle V \rangle = - \frac{1}{L_{\rm AdS}^2} \,. 
\ee
In the case at hand we see that the canonically normalized tachyon mass is $(m^2 L_{\rm AdS}^2)_{\text{tachyon}}$ $\simeq -0.998 > - 1$. 
Interestingly, the possibly generic existence of a tachyonic scalar mode within specific AdS setups was investigated in \cite{Andriot:2022brg} within the scope of the so-called AdS-TCC (or ATCC). 

In Table \ref{TABmass} we present the normalized masses of the scalars $x$, $y$ and $\tilde s^a$ as the parameter $\gamma$ takes also the values $1$ (when the seven-dimensional internal space is almost isotropic), $10^3$, $10^6$, repeating the $\gamma=10^9$ for comparison, and including also the limiting case $\gamma=2.5305 \times 10^9$. 
A negative mass eigenvalue is present in all these cases and its value is always above the Breitenlohner--Freedman bound. 
Looking at the behaviour of this mass, one could suspect that for some value of $\gamma$ around $2.5305 \times 10^9$ the BF bound may be saturated by the tachyonic scalar, giving precisely ($m^2 L_{\rm AdS}^2$)$_{\rm tachyon} = - 1$, while above such critical value of $\gamma$ supersymmetric vacua of \eqref{SuperP} within \eqref{ModuliAnsatz} cease to exist. 
However, without an analytic solution for \eqref{EqsRTS}, we can not verify this explicitly.

\begin{table} 
\begin{center}
\renewcommand{\arraystretch}{1.5}
\begin{tabular}{|c||c|}
     \hline 
     $\gamma$ & $m^2 L_{\rm AdS}^2$   \\ 
     \hline \hline
     $1$   &  $\{49.254, 7.581, 5.435, 5.435, 5.435, 5.435, 2.405, -0.9799 \}$  \\
     \hline
     $10^{3}$  &  $\{49.19188, 7.99867, 7.99467, 7.99467, 3.93721, 2.09572, 1.78518, -0.997995\}$  \\
     \hline
     $10^{6}$  &  $\{49.19179, 8, 7.99999, 7.99999, 3.93391, 2.09457, 1.77779, -0.998051\}$  \\
     \hline
     $10^{9}$ & $\{49.19179, 8, 8, 8, 3.93391, 2.09457, 1.77778, -0.998051\}$ \\
     \hline
     $2.5305 \times 10^9$ & $\{49.19179, 8, 8, 8, 3.93391, 2.09457, 1.77778, -0.998051 \}$\\
     \hline
\end{tabular}
\caption{\label{TABmass} This table exhibits the (rounded-up numerical values of the) masses of the closed string moduli as the parameter $\gamma$ takes the values $1$, $10^3$, $10^6$, $10^9$ and $2.5305 \times 10^9$. The negative eigenvalue remains always slightly above the BF bound.}
\end{center}
\end{table}

\section{Swampland considerations} 
\label{Swampland}

As we have seen by analysing the mass spectrum of the model under consideration, our setup provides a realization of the conjectured negative mass of \cite{Andriot:2022brg}. 
We can now discuss some further interesting aspects of the Swampland Program within our construction, related (in particular) to the Distance Conjecture and the Spin-2 Conjecture.

\subsection{The Distance Conjecture}  

Clearly, it is interesting to study the interpolation between the vacua that are characterized by full scale separation and those ones that have a large radius, and see how the KK-modes behave. 
This can be done by following the setup of \cite{Shiu:2022oti,Shiu:2023bay}\footnote{For another perspective on the evaluation of the distance see also \cite{Li:2023gtt}.},
varying the value of $N$ with the use of D4-branes while keeping $\gamma$ fixed. 
Specifically, for appropriate (and fixed) values of $\gamma$ and when $N$ is not parametrically large, we have a large value of $r_7$; by increasing $N$, we appreciate that full scale separation takes place, while the tower of KK-modes has, as we will see, an exponentially dropping mass to realize the Distance Conjecture. 

To be precise, let us remind the reader that we have 
\be
F_4 = (2 \pi)^3 N 
\left[
\left( 
\Psi_1 + \Psi_2 + \Psi_3 - 3 \Psi_4 
\right) 
+ \gamma 
\left(
\Psi_5 + \Psi_6 - 2 \Psi_7 
\right)  
\right]
\,, 
\ee
and $H_3= (2 \pi)^2 K \sum_i \Phi_i$. 
Similarly to \cite{Shiu:2022oti}, where it is shown that with the inclusion of appropriate D4-branes the value of $N$ can change at the cost of introducing a new scalar in the moduli space, let us consider $\tilde {\rm N}_i$ D4-branes wrapping the three-dimensional external space and a 2-cycle within an internal 3-cycle $\Sigma_{3,i}$\footnote{We will not discuss here the scaling of the potential to justify the use of such D4-branes within the effective description: it is fairly accepted at this point that D4-branes are admitted and facilitate the flux jump. In any case we do not have more convincing or newer arguments to offer on top of the ones presented in \cite{Shiu:2022oti}.}. 
We furthermore parametrize the metric as 
\be
{\rm d}s^2_{10} = e^{2\alpha v} {\rm d}s^2_3 + s_i^{\frac23} \left[ {\rm d}\psi_i^2 + w_i^2(\psi_i) g_{mn} {\rm d}y^m {\rm d}y^n \right] + e^{2\beta v} {\rm d}s^2_{4,i} \,,
\ee
where $\psi_i$ is the transverse coordinate to the 2-cycle along ($y^m$, $y^n$), which the D4-brane occupies within $\Sigma_{3,i}$. 
Here the real scalar $v$ is a convenient rewriting of the volume modulus with 
\be
e^{\beta v} = {\text{Vol}}(X_7)^{1/7} \,, \quad e^{\alpha v} = {\text{Vol}}(X_7)^{-1} \,, 
\ee
where the constants $\alpha$ and $\beta$ take the values $\alpha=\sqrt{7}/4$ and $\beta=-1/(4 \sqrt{7})$.

The full D4-brane action in the Einstein frame then takes the form 
\be
S_{{\rm D4},i} = S^{\textrm{DBI}}_{\textrm{D4},i} + S^{\textrm{CS}}_{\textrm{D4},i} \,
\ee
with
\begin{equation}
\begin{array}{lll}
S^{\textrm{DBI}}_{\textrm{D4},i} &=& - |\tilde {\rm N}_i| T_{4} \displaystyle\int {\rm d}^{5}\xi \, e^{\frac{\phi}{4}} \sqrt{-\textrm{det}\left[g_{AB} + e^{-\frac{\phi}{2}} (2 \pi \, F_{AB} - B_{AB}) \right]} \ , \\[4mm]
S^{\textrm{CS}}_{\textrm{D4},i} &=& |\tilde {\rm N}_i| T_{4} \displaystyle\int \left. \left(  \sum_r C_{r} \wedge e^{-B_{2}}\right) \wedge   e^{2 \pi F} \, \right|_{5}  \,,
\end{array}
\end{equation}
the tension of a D4-brane being $T_4 = \frac{1}{(2\pi)^4}$ (as its charge), once $\alpha'=1$. 
In the absence of world-volume flux and expressing the $B_2$ field as
\be
B_{2, i}= [B_2]_{m n, i} (\psi_i) {\rm d}y^m\wedge{\rm d}y^n = (2\pi)^2 U_i(\psi_i) \epsilon_{m n} \sqrt{{\rm det}[g_{pq}]}{\rm d}y^m\wedge{\rm d}y^n \,,
\ee
which has to satisfy $\int_{\Sigma_{3,i}} {\rm d}B_{2,i} = h = (2\pi)^2 K$ and for which $U_i({\rm sgn}({\tilde N}_i)\pi f) = {\rm sgn}({\tilde N}_i) f h$ (in order to resolve the anomaly due to the potential appearance of a net D2 charge), we get
\be
\begin{aligned}
S_{\rm{D4},i} &= -\frac{(2\pi)^{21}|\tilde {\rm N}_i| T_4}{8} \int_{{\rm AdS}_3} {\rm d}^3 x \sqrt{-{\rm det}[g^{(3)}_{\mu\nu}]} \int_{\rm 2-cycle} {\rm d}^2 y \sqrt{{\rm det}[g^{(2)}_{mn}]} \times \\[2mm] & \qquad \qquad \qquad  \quad \times e^{\frac{\phi}{4}} e^{3\alpha v} \sqrt{1 + \frac{4 s_i^{\frac23} e^{-2\alpha v}}{(2\pi)^{14}} (\partial \psi_i)^2} \, \sqrt{s_i^{\frac43} w_i(\psi_i)^4 + e^{-\phi} (2\pi)^4 U_i(\psi_i)^2} + \dots \,,
\end{aligned}
\ee
or
\be
\begin{aligned}
S_{\rm{D4},i} &= - \frac{(2\pi)^{21}|\tilde {\rm N}_i| T_4 v_{2,i}}{8} \int_{{\rm AdS}_3} {\rm d}^3 x \sqrt{-{\rm det}[g^{(3)}_{\mu\nu}]} \times \\[2mm]& \qquad \qquad \qquad \qquad \ \times e^{\frac{\phi}{4}} e^{3\alpha v} \sqrt{1 + \frac{4 s_i^{\frac23} e^{-2\alpha v}}{(2\pi)^{14}} (\partial \psi_i)^2} \, \sqrt{s_i^{\frac43} w_i(\psi_i)^4 + e^{-\phi} (2\pi)^4 U_i(\psi_i)^2} + \dots \,,
\end{aligned}
\ee
where we have rescaled the three-dimensional metric $g^{(3)}_{\mu\nu}$ as  $g^{(3)}_{\mu\nu} \rightarrow \frac{(2\pi)^{14}}{4} g^{(3)}_{\mu\nu}$ to make contact with the standard conventions for three-dimensional supergravity theories, and the ``dots'' stand for terms that do not play a role for the upcoming discussion. 
From the expression above we can extract the field space metric of the position scalar $\psi_i$ that is 
\be
\begin{aligned}
g_{\psi_i\psi_i} &= \frac{(2\pi)^{7}|\tilde {\rm N}_i| T_4 v_{2,i}}{2} \, e^{\frac{\phi}{4}-7\beta v} s_i^{\frac23} \sqrt{s_i^{\frac43} w_i(\psi_i)^4 + e^{-\phi} (2\pi)^4 U_i(\psi_i)^2} = \\& = \lambda e^{\frac{\phi}{4}-3\beta v} {\tilde s}_i^{\frac43} \sqrt{w_i(\psi_i)^4 + {\tilde s}_i^{-\frac43} e^{-\phi-4\beta v} (2\pi)^4 U_i(\psi_i)^2} \,.
\end{aligned}
\ee
Other than $g_{\psi_i\psi_i}$, in order to evaluate the distance between the vacuum where a radius (e.g. $r_7$) is large and the one where full scale separation is realized, we need the metric components $g_{\phi\phi}$, $g_{vv}$ and $g_{{\tilde s}^a {\tilde s}^b}$, which can be extracted from the dimensionally reduced ten-dimensional action over the G2 space, once the fields $\phi$, $v$ and $\tilde{s}^a$ have been canonically normalized:
\be
g_{\phi\phi} = \frac12 \,, \quad 
g_{vv} = \frac12 \quad \text{and} \quad 
g_{{\tilde s}^a {\tilde s}^b} = \frac{1+\delta_{ab}}{2{\tilde s}^a {\tilde s}^b} \quad \text{for } a, b= 1,...,6 \,. 
\ee
The field space distance of interest is
\be
\Delta = \int_{\zeta=0}^{\zeta=1} {\rm d}\zeta \sqrt{g_{{\cal I} {\cal J}} \frac{{\rm d}{\varphi}^{{\cal I}}}{{\rm d}\zeta}\frac{{\rm d}{\varphi}^{{\cal J}}}{{\rm d}\zeta}} \,,
\ee
where the indices ${\cal I}, {\cal J}$ run over the fields $\varphi = \{\psi_i, \phi, v, {\tilde s}^a\}$ and $g_{{\cal I}, {\cal J}}$ denotes the corresponding metric element. By exploiting reparametrization invariance we can impose that $\zeta=0$ corresponds to the circumstance where the radius $r_7$ is large and $\zeta=1$ is associated with the realization of full scale separation.
When considering the contributions from the relevant 3-cycles $\Sigma_{3,i} \equiv \Phi_i$ for $i = 1,...7$, and redefining $\sigma$, $\rho$ and $\tau$ as
\be
\sigma = e^{3s} \,, \quad \rho = e^{3r} \,, \quad \tau = e^{3t} \,, 
\ee
we get
\be
\begin{aligned}
\Delta &= \int_{0}^{1} {\rm d}\zeta \Bigg[  \sum_{i=1}^7 \lambda e^{\frac{\phi}{4}-3\beta v} {\tilde s}_i^{\frac43} \sqrt{w_i(\psi_i)^4 + {\tilde s}_i^{-\frac43} e^{-\phi-4\beta v} (2\pi)^4 U_i(\psi_i)^2} \left(\frac{{\rm d}\psi_i}{{\rm d}\zeta}\right)^2 + \frac12 \left(\frac{{\rm d}\phi}{{\rm d}\zeta}\right)^2  + \\& + \frac12 \left(\frac{{\rm d}v}{{\rm d}\zeta}\right)^2   + 54 \left(\frac{{\rm d}s}{{\rm d}\zeta}\right)^2 + 9 \left(\frac{{\rm d}r}{{\rm d}\zeta}\right)^2 + 27 \left(\frac{{\rm d}t}{{\rm d}\zeta}\right)^2 + 27 \frac{{\rm d}s}{{\rm d}\zeta}\frac{{\rm d}r}{{\rm d}\zeta} + 54 \frac{{\rm d}s}{{\rm d}\zeta}\frac{{\rm d}t}{{\rm d}\zeta} + 18 \frac{{\rm d}r}{{\rm d}\zeta}\frac{{\rm d}t}{{\rm d}\zeta}  \Bigg]^{\frac12} \,.
\end{aligned}
\ee
Keeping in mind that we are simply interested in extracting how $\Delta$ scales with $N$ and considering the numerical results of Subsection \ref{Numerics}, let us massage the previous expression by grouping the 3-cycles $\Sigma_{3,i}$ according to their approximate size: in particular, we are going to set $s_1 = s_2 = s_3 = e^{3\beta v} \sigma = e^{3\beta v + 3s}$ and $s_4 = e^{3\beta v} \rho = e^{3\beta v + 3r} \sim e^{3\beta v + 3t} = e^{3\beta v} \tau = s_5 = s_6 \sim s_7$; we will denote as ${\hat s}_1$ and ${\hat s}_2$ the typical sizes of these two classes of 3-cycles. Moreover, we will refer to the position scalars corresponding to such two groups as ${\hat \psi}_1$ and ${\hat \psi}_2$, respectively. Then, if $r \sim t \simeq u$,
\be
\begin{aligned}
\Delta \approx \int_{0}^{1} {\rm d}\zeta \Bigg[ & \lambda e^{\frac{\phi}{4}-3\beta v} \Bigg( 3 e^{4s} \sqrt{{\hat w}_1({\hat \psi}_1)^4 + e^{-\phi-4\beta v-4s} (2\pi)^4  {\hat U}_1({\hat \psi}_1)^2} \left(\frac{{\rm d}{\hat \psi}_1}{{\rm d}\zeta}\right)^2 + \\& + 4 e^{4u} \sqrt{{\hat w}_2({\hat \psi}_2)^4 + e^{-\phi-4\beta v-4u} (2\pi)^4 {\hat U}_2({\hat \psi}_2)^2} \left(\frac{{\rm d}{\hat \psi}_2}{{\rm d}\zeta}\right)^2 \Bigg) + \\& + \frac12 \left(\frac{{\rm d}\phi}{{\rm d}\zeta}\right)^2 + \frac12 \left(\frac{{\rm d}v}{{\rm d}\zeta}\right)^2 + 54 \left(\frac{{\rm d}s}{{\rm d}\zeta}\right)^2 + 54 \left(\frac{{\rm d}u}{{\rm d}\zeta}\right)^2 + 81 \frac{{\rm d}s}{{\rm d}\zeta}\frac{{\rm d}u}{{\rm d}\zeta} \Bigg]^{\frac12} \,.
\end{aligned}
\ee
Let us importantly observe that, since the functions $w_i(\psi_i)$ and $U_i(\psi_i)$ do not scale with $N$, and because of the scalings with $N$ of $e^{\phi}$ and $e^{\beta v}$, 
\be
\sqrt{{\hat w}_k({\hat \psi}_k)^4 + e^{-\phi-4\beta v} {\tilde {\hat s}_k}^{-\frac43} (2\pi)^4 {\hat U}_k({\hat \psi}_k)^2} \approx {\hat w}_k({\hat \psi}_k)^2 \,,
\ee
the factor $e^{-\phi-4\beta v}$, which carries the dependence on $N$ and accompanies ${\hat U}_k({\hat \psi}_k)^2$, being $\mathcal{O}(N^{-1})$, i.e. consistently suppressed for large $N$. As a consequence, we obtain
\be \label{SimplifiedDelta}
\begin{aligned}
\Delta \approx \int_{0}^{1} {\rm d}\zeta \Bigg[ & e^{\frac{\phi}{4}-3\beta v+4s} \left(\frac{{\rm d}{\hat \chi}_1}{{\rm d}\zeta}\right)^2 +  e^{\frac{\phi}{4}-3\beta v+4u} \left(\frac{{\rm d}{\hat \chi}_2}{{\rm d}\zeta}\right)^2 + \\& + \frac12 \left(\frac{{\rm d}\phi}{{\rm d}\zeta}\right)^2 + \frac12 \left(\frac{{\rm d}v}{{\rm d}\zeta}\right)^2 + 54 \left(\frac{{\rm d}s}{{\rm d}\zeta}\right)^2 + 54 \left(\frac{{\rm d}u}{{\rm d}\zeta}\right)^2 + 81 \frac{{\rm d}s}{{\rm d}\zeta}\frac{{\rm d}u}{{\rm d}\zeta} \Bigg]^{\frac12} \,,
\end{aligned}
\ee
where we also exploited the following field re-definitions:
\be
\frac{{\rm d}{\hat \chi}_1}{{\rm d}{\hat \psi_1}} = \sqrt{3\lambda} |{\hat w}_1({\hat \psi}_1)| \quad \text{and} \quad \frac{{\rm d}{\hat \chi}_2}{{\rm d}{\hat \psi_2}} = \sqrt{4\lambda} |{\hat w}_2({\hat \psi}_2)| \,.
\ee
In order to identify the geodesic path we have to recognize (at least approximately) the geometric structure of the scalar field space under consideration. To this end we firstly observe that the second term in the first line of \eqref{SimplifiedDelta} is highly suppressed, by $e^{4u-4s} \sim \left(\frac{\rho}{\sigma}\right)^{\frac43} \sim \left(\frac{\tau}{\sigma}\right)^{\frac43} \ll 1$, with respect to the term that precedes it in the same \eqref{SimplifiedDelta}\footnote{This, in accordance to the numerical results of Subsection \ref{Numerics} when considering, for instance, the (relevant) cases $\gamma = 10^{3}, 10^6$ or $\gamma \sim \mathcal{O}(10^9)$.}. We are therefore going to ignore such contribution from the open string moduli sector and retain only that associated with the bigger internal 3-cycles, with volumes $s_1$, $s_2$ and $s_3$. This being established, we have to perform some useful field transformations. We trade $s$ for
\be
s = {\breve s} - \frac34 u 
\ee
and we further rescale ${\breve s}$ and $u$ as
\be
{\breve s} = \frac{{\check s}}{\sqrt{108}} \quad \text{and} \quad u = \sqrt{\frac{4}{189}} {\check u} \,. 
\ee
Moreover, we can perform an $O(4)$ transformation via the orthogonal matrix\footnote{As required by the orthogonal group, the matrix $O$ is satisfies $O O^T = (\text{det}[O])^{\frac12} \mathbb{1}_4$.}
\be
O = \left(\begin{matrix}  -\frac18 & -\frac{3\beta}{2} & \frac{1}{3\sqrt{3}} & -\frac{1}{\sqrt{21}} \\ \frac{3\beta}{2} & -\frac18 & -\frac{1}{\sqrt{21}} &  -\frac{1}{3\sqrt{3}} \\ -\frac{1}{3\sqrt{3}} &  \frac{1}{\sqrt{21}} & -\frac18 & -\frac{3\beta}{2}  \\ \frac{1}{\sqrt{21}} & \frac{1}{3\sqrt{3}} & \frac{3\beta}{2} & -\frac18 \end{matrix}\right) \qquad \text{with} \quad \text{det}[O] = \left(\frac{1213+27216\beta^2}{12096}\right)^2
\ee
so that
\be
\left(\begin{matrix} \phi \\ v \\ \check s \\ \check u \end{matrix}\right) = \frac{12096}{1213+27216\beta^2} \left(\begin{matrix}  -\frac18 & -\frac{3\beta}{2} & \frac{1}{3\sqrt{3}} & -\frac{1}{\sqrt{21}} \\ \frac{3\beta}{2} & -\frac18 & -\frac{1}{\sqrt{21}} &  -\frac{1}{3\sqrt{3}} \\ -\frac{1}{3\sqrt{3}} &  \frac{1}{\sqrt{21}} & -\frac18 & -\frac{3\beta}{2}  \\ \frac{1}{\sqrt{21}} & \frac{1}{3\sqrt{3}} & \frac{3\beta}{2} & -\frac18 \end{matrix}\right)  \left(\begin{matrix} z_1 \\ z_2 \\ z_3 \\ z_4 \, \end{matrix}\right) \,. 
\ee
Then, after redefining ${\hat \psi}_1$ to be
\be
{\hat \psi}_1 = \frac{1}{\sqrt{2} ({\rm det}[O])^{\frac14}} {\hat h}_1 \,,
\ee
we get
\be
\Delta \approx \frac{1}{\sqrt{2} ({\rm det}[O])^{\frac14}} \int_{0}^{1} {\rm d}\zeta \sqrt{  e^{-2z_1} \left(\frac{{\rm d}{\hat h}_1}{{\rm d}\zeta}\right)^2  + \left(\frac{{\rm d}z_1}{{\rm d}\zeta}\right)^2 +  \left(\frac{{\rm d}z_2}{{\rm d}\zeta}\right)^2 + \left(\frac{{\rm d}z_3}{{\rm d}\zeta}\right)^2 + \left(\frac{{\rm d}z_4}{{\rm d}\zeta}\right)^2 } \,.
\ee
By further redefining $z_1$ so that ${\hat h}_2 = e^{-z_1}$ the distance $\Delta$ takes the form
\be
\Delta \approx \frac{1}{\sqrt{2} ({\rm det}[O])^{\frac14}} \int_{0}^{1} {\rm d}\zeta \sqrt{\frac{1}{{\hat h}_2^2}  \left[\left(\frac{{\rm d}{\hat h}_1}{{\rm d}\zeta}\right)^2  + \left(\frac{{\rm d}{\hat h}_2}{{\rm d}\zeta}\right)^2 \right]+  \left(\frac{{\rm d}z_2}{{\rm d}\zeta}\right)^2 + \left(\frac{{\rm d}z_3}{{\rm d}\zeta}\right)^2 + \left(\frac{{\rm d}z_4}{{\rm d}\zeta}\right)^2 } \,.
\ee
The geodesic path can thus be approximated as
\be
{\hat h}_1 = l \sin[f(\zeta)] + {\hat h}_{1,0} \,, \quad {\hat h}_2 = l \cos[f(\zeta)] 
\ee
with
\be
f(\zeta) = 2\arctan\left[\sinh\left[\frac{d_1 \zeta + d_2}{2}\right]\right]
\ee
and
\be
z_2 = d_3 \zeta + d_4 \,, \quad z_3 = d_5 \zeta + d_6 \,, \quad z_4 = d_7 \zeta + d_8 \,,
\ee
so that the geodesic distance is estimated to be
\be
\Delta \approx \frac{1}{\sqrt{2} ({\rm det}[O])^{\frac14}} \sqrt{d_1^2 + d_3^2 + d_5^2 + d_7^2} \,,
\ee
where the coefficients $\{d_j\}_{j=1,3,5,7}$ are determined by the boundary values of the fields ${\hat h}_1$, ${\hat h}_2$, $z_2$, $z_3$ and $z_4$ corresponding to the two vacuum configurations we are interpolating between, namely the one where a radius (e.g. $r_7$) is large (at $\zeta=0$) and the one where full scale separation is achieved (at $\zeta=1$).
Accounting for the scalings (up to numerical factors)
\be
{\hat \psi}_1 \sim N \,, \quad \phi \sim \log N \,, \quad v \sim \log N
\ee
we can deduce, again up to numerical factors, the behaviours
\be
{\hat h}_1 \sim N \sim {\hat h}_2 \,, \quad z_k \sim \log N \quad \text{for } k=2,3,4 
\ee
so that
\be
d_j \sim \log N  \quad  \text{for } j=1,3,5,7 \,, 
\ee
and eventually 
\be
\Delta \sim \delta \log N \,,
\ee
$\delta$ being a numerical coefficient that can be crudely estimated to be $\delta \simeq 24.6$. 
The crucial result here is the logarithmic behaviour, which is precisely the way $\Delta$ should scale with $N$, if the Distance Conjecture
\be
\frac{m_{\rm KK}(\zeta=1)}{m_{\rm KK}(\zeta=0)} \sim e^{- \kappa \Delta} \,, \quad \kappa \sim \mathcal{O}(1) \, 
\ee
is realized. 
This happens because we already have 
\be
\frac{m_{\rm KK}(\zeta=1)}{m_{\rm KK}(\zeta=0)} \sim N^{-\frac72} \,,
\ee
as one can easily infer from \eqref{KKmass} accounting for the scaling with $N$ of the volume ${\rm Vol}(X_7)$ and the radii $r_i$.

We conclude that for the highly anisotropic vacua under analysis the Distance Conjecture is realized, always with the inclusion of the appropriate D4-branes. 
This distance is along the path that connects AdS$_3\times$S$^1$ and fully scale-separated AdS$_3$, 
with the tower of the $r_7$ KK-modes turning full scale separation {\it on} and {\it off}.

\subsection{The Strong Spin-2 Conjecture}

A further interesting connection of our work with the Swampland Program has to do with the Strong Spin-2 Conjecture \cite{Klaewer:2018yxi}, which states that in any theory with massive spin-2 fields there has to be a tower of states related to the massive spin-2 modes even in the absence of a massless spin-2 excitation. 
Indeed, since standard gravity in three dimensions does not have local excitations (i.e. there are no massless gravitons), the only spin-2 states are the ones related to the KK-modes of the graviton. 
Since they are KK-modes, these states essentially form a tower that has as characteristic mass scale the mass of the lowest one, which also defines the cut-off of the three-dimensional effective theory $\Lambda_{\rm UV}\sim m_{\rm spin-2}$. 
In AdS space having such low cut-off is not an inconsistency per se, if one is protected by supersymmetry; it just means that the system is not inherently three-dimensional. 
In this way our construction provides a non-trivial example in favour of \cite{Klaewer:2018yxi}. 

This observation also tells us that, even though we have been able to find AdS$_3\times$S$^1$ solutions, we would not be able to find dS$_3\times$S$^1$ solutions within a controlled setup. 
This happens because, if we were able to have dS$_3$ and to lower the KK-modes such that $m_{\rm spin-2} \ll H$, then we would also have $\Lambda_{\rm UV} \ll H$ which signals an inherent inconsistency for de Sitter. 
Actually, such an inconsistency can be directly deduced from the Higuchi bound \cite{Higuchi:1986py},
which in three dimensions is 
\be 
\nn 
m_{\rm spin-2}^2 \geq H^2 \,. 
\ee
Therefore, one can never make a single radius parametrically grow in de Sitter and break scale separation as we did here for the AdS case.

\section{Discussion}

In this work we have studied classical flux vacua of massive type-IIA supergravity with appropriate numbers of branes and orientifold planes. 
All the constructions that we have analyzed are characterized by proper flux quantization, closed string moduli stabilization, large volume and weak coupling, while the O6-planes are smeared. 
In such setup the compact space is a seven-dimensional G2 toroidal orbifold and the non-compact space is AdS$_3$. 
Our aim was to find configurations that induce a considerable anisotropy in the compact space such that one dimension can become large and comparable to the non-compact space length scale, whereas the other six dimensions become considerably smaller than the non-compact space length scale. 
In this way the original ten-dimensional space reduces to a product of a large four-dimensional external one and a small six-dimensional one, with (non-parametric) scale separation between the two. 
When, instead, the $F_4$ flux takes parametrically large values we recover AdS$_3$ parametrically scale-separated from the seven-dimensional internal space, while preserving (as $\gamma$ is fixed) the amount of anisotropy between one of the internal dimensions (e.g. $L_{{\rm KK},7}$) and the remaining six ones (i..e. $L_{{\rm KK},a}$). 
This behaviour can be schematically described by the figure below for some $F_4$ flux N (N' being the value that allows one large radius): 
\\[-0.8cm]
\begin{center}
\includegraphics[scale=0.5]{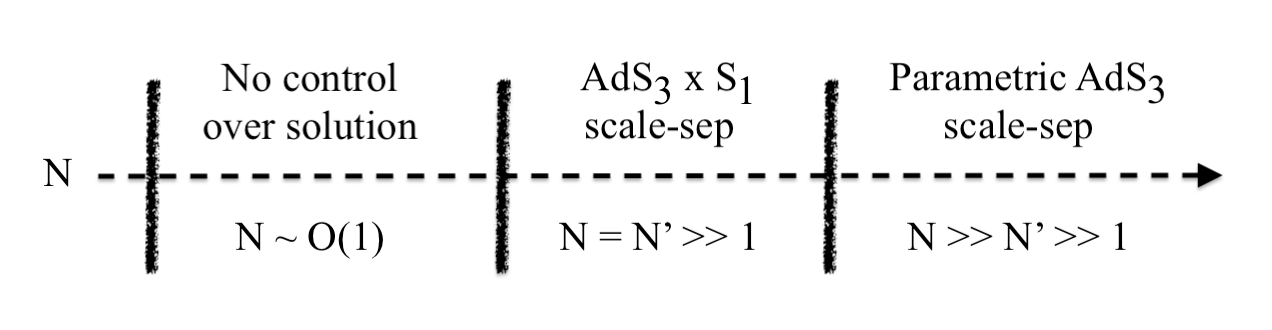} 
\end{center}
{\ }
\\[-1.2cm] 
This happens because such vacua have the interesting property that, while $N$ increases, both $L_{{\rm KK},i}$ and $L_{\rm AdS}$ increase; 
however, their ratio changes proportionally to $L_{{\rm KK},i}^2 / L_{\rm AdS}^2 \sim 1/N$, thus always restoring full scale separation at parametrically large $N$.

Our work leaves a series of open questions for future investigation. 
Firstly, a full analysis of all types of flux vacua that can be constructed from the G2 orientifold of \cite{Farakos:2020phe} is missing. 
The full list of choices that one can make for the $F_4$ and the $H_3$ flux are still not classified; any development in that direction would be welcome. 
Secondly, in all of these constructions the open string moduli are assumed to be stabilized. 
However, a careful analysis that checks this and clarifies the full moduli space and superpotential would be interesting. 
Thirdly, we are still working with a toroidal orbifold here; 
the explicit study of different G2 spaces or G2-structure manifolds is still an open question 
(however, see \cite{Passias:2020ubv} for general conditions on the G2-structure from type-IIB), 
and the only explicit development providing a superpotential in this direction was the analysis of type-IIB supergravity on co-calibrated G2 toroidal orbifolds with O5/O9-planes \cite{Emelin:2021gzx}, while a general superpotential for such type-II constructions was derived in \cite{VanHemelryck:2022ynr}. 
Finally, as far as the more fundamental issues are concerned, one could try to go to higher orders in the O6-plane backreaction since, for the moment, only the leading order was analyzed in \cite{Emelin:2022cac}.

\section*{Acknowledgements} 

FF thanks the Physics Division of NTUA for the hospitality during the completion of this work. MM is partially supported by the Ministry of Science and Universities through the Spanish grant MCIU-22-PID2021-123021NB-I00 and by FICYT through the Asturian grant SV-PA-21-AYUD/2021/52177.

\appendix

\section{Small $\gamma$ values with different anisotropy}

As a complement to what we discussed in the bulk of the paper, it is interesting to investigate what happens when $\gamma$ takes small values. 
For the sake of the presentation, we will focus on two representative values of $\gamma$, namely $\gamma = 10^{-3}$ and $\gamma = 5 \times 10^{-6}$ (being conscious of the fact that for $\gamma \sim 4 \times 10^{-6}$ moduli stabilization of \eqref{SuperP} within the ansatz \eqref{ModuliAnsatz} ceases to exist, and that, since $q= \gamma f$, $\gamma N \in \mathbb{Z}$, which is satisfied in our examples.). 

The equations in \eqref{EqsRTS} are solved by the values of $\sigma$, $\rho$ and $\tau$ that are reported in Table \ref{TAB8}; as a consequence, we also get that the string coupling is given by
\be
g_s \approx 71.33 \times \frac{|K|}{N^{3/4} |M|^{1/4}} \,, \quad \text{for } \gamma = 10^{-3} 
\ee
or
\be
g_s \approx 1.0085 \times 10^3\times \frac{|K|}{N^{3/4} |M|^{1/4}} \,, \quad  \text{for } \gamma = 5 \times 10^{-6} \,,
\ee
and that the radii take the values that Table \ref{TAB9} exhibits. We see once again that, after fixing $K$ and $M$ consistently with the tadpole conditions (e.g. $K = -16$ and $M = -1$ ), we can end up in the desired weak string coupling and large internal volume regime by taking a sufficient amount of $N$ units of the $F_4$ flux.

Besides the possibility to recover full scale separation by taking $N$ parametrically large, we would like to understand once more if for some moderate (but still sufficiently large) value of $N$ we could effectively obtain a four-dimensional external space, AdS$_3 \times S^1$, scale-separated from a six-dimensional internal one. 
This means that, in light of our previous discussion and the results presented in Table \ref{TAB9}, we want to investigate whether
\be
\frac{L_{\rm{KK},5}^2}{L_{\rm AdS}^2} = \frac{4}{\pi^2} \left(\frac{P}{(2\pi)^7} \right)^2 (\text{Vol}(X_7))^2 \, r_5^2 \gtrsim  1
\,, \quad 
\frac{L_{\rm{KK},i\neq5}^2}{L_{\rm AdS}^2} \ll 1 
\,. 
\ee 
More precisely, we have the ratios 
\be
\frac{L_{\rm{KK},5}^2}{L_{\rm AdS}^2} \Big{|}_{\gamma = 10^{-3}} \approx  9 \times 10^4 \times \frac{K^2 |M|}{N} 
\,, \quad 
\frac{L_{\rm{KK},5}^2}{L_{\rm AdS}^2} \Big{|}_{\gamma = 5 \times 10^{-6}} \approx  3.6 \times 10^{9} \times \frac{K^2 |M|}{N} \,, 
\ee
which offer quite a wide range of values of $N$ that can be checked. 
Indeed, when fixing $K = -16$ and $M = -1$, we find, for $\gamma = 10^{-3}$, the results that are presented in Table \ref{TAB10}, and, for $\gamma = 5 \times 10^{-6}$, we end up with the values that are reported in Table \ref{TAB11}, together with the relative behaviour of the string coupling and the other radii.

\begin{table}
\begin{center}
\renewcommand{\arraystretch}{1.5}
\begin{tabular}{|c||c||c||c|}
     \hline 
     $\gamma$ & $\sigma$ & $\rho$ & $\tau$  \\ 
     \hline \hline
     $10^{-3}$   &  $0.212$  & $0.033$ & $268.4$ \\
     \hline
     $5 \times 10^{-6}$  &  $0.047$ & $0.007$ & $11811.8$ \\
     \hline
\end{tabular}
\caption{
\label{TAB8} 
This table presents the critical values of $\sigma$, $\rho$ and $\tau$ as $\gamma$ takes the values $\gamma = 10^{-3}$, $5 \times 10^{-6}$. 
}
\end{center}
\end{table}

\begin{table}
\begin{center}
\renewcommand{\arraystretch}{1.5}
\begin{tabular}{|c||c||c||c||c||c|}
     \hline 
     $\gamma$ & $\frac{r_{1,3}}{\frac{2\pi N^{7/16}}{|M|^{3/16} |K|^{1/4}}}$ & $\frac{r_{2,4}}{\frac{2\pi N^{7/16}}{|M|^{3/16} |K|^{1/4}}}$ & $\frac{r_{5}}{\frac{2\pi N^{7/16}}{|M|^{3/16} |K|^{1/4}}}$ & $\frac{r_{6}}{\frac{2\pi N^{7/16}}{|M|^{3/16} |K|^{1/4}}}$ & $\frac{r_{7}}{\frac{2\pi N^{7/16}}{|M|^{3/16} |K|^{1/4}}}$  \\ 
     \hline \hline
     $10^{-3}$   &  $0.402$  & $0.469$ & $36.37$ & $0.005$ & $0.029$ \\
     \hline
     $5 \times 10^{-6}$  & $0.207$  & $0.242$ & $265.2$ & $0.0002$ & $0.001$ \\
     \hline
\end{tabular}
\caption{\label{TAB9} This table shows the values of the radii $r_i$ as $\gamma$ takes the values $\gamma = 10^{-3}$, $5 \times 10^{-6}$. One can clearly notice that, as $\gamma$ becomes smaller, $r_5$ becomes bigger than the other six radii $r_a$ (with $a\neq5$) and $r_7$, for fixed flux units.}
\end{center}
\end{table}

\begin{table}
\begin{center}
\renewcommand{\arraystretch}{1.5}
\begin{tabular}{|c||c||c||c||c||c||c||c||c|}
     \hline 
     $$ &  $N$ & $g_s$ & $r_{i<5}$ & $r_{i>5}$ & $r_5$ & $\frac{L_{\rm{KK},5}^2}{L_{\rm AdS}^2}$ & $\frac{L_{\rm{KK},i<5}^2}{L_{\rm AdS}^2}$ & $\frac{L_{\rm{KK},i>5}^2}{L_{\rm AdS}^2}$ \\ 
     \hline \hline
     $(a)$  & $10^{10}$ & ${\cal O}(10^{-5})$ & ${\cal O}(10^4)$ & ${\cal O}(10^{2\div3})$  &  $2.7\times10^6$ & $2.3\times10^{-3}$ & $ {\cal O}(10^{-7})$ & ${\cal O}(10^{-11})$  \\
     \hline
     $(b)$  & $10^7$ & ${\cal O}(10^{-3})$ & ${\cal O}(10^3)$ & ${\cal O}(10^{1\div2})$ & $1.3\times 10^5$ &  $2.3$ & ${\cal O}(10^{-4})$ & ${\cal O}(10^{-8})$  \\
     \hline
     $(c)$  & $10^5$ & ${\cal O}(10^{-1})$ & $ {\cal O}(10^2) $& ${\cal O}(10^{0\div1})$ & $1.8\times10^4$ & $2.3 \times 10^2$ & $ {\cal O}(10^{-2})$ & $ {\cal O}(10^{-6})$ \\
     \hline
\end{tabular}
\caption{\label{TAB10} This table shows the three interesting regimes one can end up with while changing $N$ for $\gamma = 10^{-3}$, once the other flux units have been fixed, namely $K=-16$ and $M=-1$. 
When a circumstance like (a) realizes, one has full scale separation; 
if, instead, one works with cases similar to (b) or (c), then the radius $r_5$ disentangles from the other six radii and the external space becomes effectively AdS$_3\times$S$^1$. 
For the in-between values of $N$ one gets of course intermediate results. 
Note that $N$ can not be too small in order for the large volume/weak coupling condition to still be satisfied.} 
\end{center}
\end{table}

\begin{table}
\begin{center}
\renewcommand{\arraystretch}{1.5}
\begin{tabular}{|c||c||c||c||c||c||c||c||c|}
     \hline 
     $$ &  $N$ & $g_s$ & $r_{i<5}$ & $r_{i>5}$ & $r_5$ & $\frac{L_{\rm{KK},5}^2}{L_{\rm AdS}^2}$ & $\frac{L_{\rm{KK},i<5}^2}{L_{\rm AdS}^2}$ & $\frac{L_{\rm{KK},i>5}^2}{L_{\rm AdS}^2}$ \\ 
     \hline \hline
     $(a)$  & $10^{18}$ & ${\cal O}(10^{-10})$ & ${\cal O}(10^7)$ & ${\cal O}(10^{4\div5})$  &  $6.3\times10^{10}$ & $9.2\times10^{-7}$ & $ {\cal O}(10^{-14})$ & ${\cal O}(10^{-19})$  \\
     \hline
     $(b)$  & $10^{12}$ & ${\cal O}(10^{-5})$ & ${\cal O}(10^5)$ & ${\cal O}(10^2)$ & $1.5\times 10^8$ &  $0.92$ & ${\cal O}(10^{-7})$ & ${\cal O}(10^{-12})$  \\
     \hline
     $(c)$  & $10^6$ & ${\cal O}(10^{-1})$ & $ {\cal O}(10^2) $& ${\cal O}(10^{-1\div0})$ & $3.5\times10^5$ & $9.2\times 10^5$ & $ {\cal O}(10^{-2})$ & $ {\cal O}(10^{-7})$ \\
     \hline
\end{tabular}
\caption{\label{TAB11} This table shows the three interesting regimes one can end up with while changing $N$ for $\gamma = 5 \times 10^{-6}$, once the other flux units have been fixed, namely $K=-16$ and $M=-1$. 
One can again see a similar behaviour with respect to the case $\gamma=10^{-3}$, which is presented in Table \ref{TAB10}.} 
\end{center}
\end{table}

\begin{table} 
\begin{center}
\renewcommand{\arraystretch}{1.5}
\begin{tabular}{|c||c|}
     \hline 
     $\gamma$ & $m^2 L_{\rm AdS}^2$   \\ 
     \hline \hline
     $10^{-3}$   &  $\{ 49.4952, 7.9987, 7.996, 5.8683, 3.6456, 3.6456, 2.3506, -0.999965 \}$  \\
     \hline
     $5 \times 10^{-6}$  &  $\{ 49.4958, 8, 8, 5.8683, 3.6428, 3.6428, 2.3504, -0.99996\}$  \\
     \hline
\end{tabular}
\caption{\label{TABmass2} This table exhibits the (rounded-up numerical values of the) masses of the closed string moduli as the parameter $\gamma$ takes the values $10^{-3}$ and $5 \times 10^{-6}$. The negative eigenvalue remains always slightly above the BF bound.}
\end{center}
\end{table}

Taking now also into account the behaviour of the system for $\gamma > 1$, we can observe that for given $K$, $M$ and $N$, as we vary the value of $\gamma$ from $\gamma \ll 1$ to $\gamma \gg 1$, the length of the radius $r_5$ decreases while the radius $r_7$ increases in magnitude. 
As a result, if we start having scale-separated AdS$_3\times$S$^1(r_5)$ with $r_5 \gg r_7$, by increasing $\gamma$ we pass through a regime where $r_5 \sim r_7$ and end up into the regime where $r_5 \ll r_7$  with scale-separated AdS$_3\times$S$^1(r_7)$. 
This means that, while we are varying $\gamma$, there is a tower of KK-modes becoming light and simultaneously another tower of KK-modes becoming heavy. 
Here, we are not going to precisely discuss a mechanism that allows to vary $\gamma$ and realize the Distance Conjecture, but it can be done probably with actual jumping fluxes, in analogy with \cite{Basile:2023rvm}, or with the use of D4-branes similarly to \cite{Shiu:2022oti}, and actually following the same steps as in Section \ref{Swampland}.

As we did in Section \ref{MassSpectrum}, we can also evaluate the normalized masses for the moduli $x$, $y$ and $\tilde s^a$. Their numerical values are reported in Table \ref{TABmass2}, where we again notice the presence of a negative mass mode that is consistent with the BF bound.


\begin{thebibliography}{99}


\bibitem{Kachru:2003aw}
S.~Kachru, R.~Kallosh, A.~D.~Linde and S.~P.~Trivedi,
``De Sitter vacua in string theory'',
Phys. Rev. D \textbf{68} (2003), 046005
[arXiv:hep-th/0301240 [hep-th]].



\bibitem{DeWolfe:2005uu}
O.~DeWolfe, A.~Giryavets, S.~Kachru and W.~Taylor,
``Type IIA moduli stabilization'', JHEP \textbf{07} (2005), 066
[arXiv:hep-th/0505160 [hep-th]].


\bibitem{Behrndt:2004mj}
K.~Behrndt and M.~Cvetic,
``General N=1 supersymmetric fluxes in massive type IIA string theory'',
Nucl. Phys. B \textbf{708} (2005), 45-71
[arXiv:hep-th/0407263 [hep-th]].


\bibitem{Derendinger:2004jn}
J.~P.~Derendinger, C.~Kounnas, P.~M.~Petropoulos and F.~Zwirner,
``Superpotentials in IIA compactifications with general fluxes'',
Nucl. Phys. B \textbf{715} (2005), 211-233
[arXiv:hep-th/0411276 [hep-th]].


\bibitem{Lust:2004ig}
D.~Lust and D.~Tsimpis,
``Supersymmetric AdS(4) compactifications of IIA supergravity'',
JHEP \textbf{02} (2005), 027
[arXiv:hep-th/0412250 [hep-th]].


\bibitem{Camara:2005dc}
P.~G.~Camara, A.~Font and L.~E.~Ibanez,
``Fluxes, moduli fixing and MSSM-like vacua in a simple IIA orientifold'',
JHEP \textbf{09} (2005), 013
[arXiv:hep-th/0506066 [hep-th]].


\bibitem{Caviezel:2008ik}
C.~Caviezel, P.~Koerber, S.~Kors, D.~Lust, D.~Tsimpis and M.~Zagermann,
``The Effective theory of type IIA AdS(4) compactifications on nilmanifolds and cosets'',
Class. Quant. Grav. \textbf{26} (2009), 025014
[arXiv:0806.3458 [hep-th]].


\bibitem{Tsimpis:2012tu}
D.~Tsimpis,
``Supersymmetric AdS vacua and separation of scales'',
JHEP \textbf{08} (2012), 142
[arXiv:1206.5900 [hep-th]].


\bibitem{Gautason:2015tig}
F.~F.~Gautason, M.~Schillo, T.~Van Riet and M.~Williams,
``Remarks on scale separation in flux vacua'',
JHEP \textbf{03} (2016), 061
[arXiv:1512.00457 [hep-th]].


\bibitem{Lust:2020npd}
D.~L\"ust and D.~Tsimpis,
``AdS$_{2}$ type-IIA solutions and scale separation'',
JHEP \textbf{07} (2020), 060
[arXiv:2004.07582 [hep-th]].


\bibitem{Banks:2006hg}
T.~Banks and K.~van den Broek,
``Massive IIA flux compactifications and U-dualities'',
JHEP \textbf{03} (2007), 068
[arXiv:hep-th/0611185 [hep-th]].


\bibitem{McOrist:2012yc}
J.~McOrist and S.~Sethi,
``M-theory and Type IIA Flux Compactifications'', 
JHEP \textbf{12} (2012), 122
[arXiv:1208.0261 [hep-th]].


\bibitem{Lust:2019zwm}
D.~L\"ust, E.~Palti and C.~Vafa,
``AdS and the Swampland'',
Phys. Lett. B \textbf{797} (2019), 134867
[arXiv:1906.05225 [hep-th]].


\bibitem{Acharya:2006ne}
B.~S.~Acharya, F.~Benini and R.~Valandro,
``Fixing moduli in exact type IIA flux vacua'',
JHEP \textbf{02} (2007), 018
[arXiv:hep-th/0607223 [hep-th]].


\bibitem{Blaback:2010sj}
J.~Blaback, U.~H.~Danielsson, D.~Junghans, T.~Van Riet, T.~Wrase and M.~Zagermann,
``Smeared versus localised sources in flux compactifications'',
JHEP \textbf{12} (2010), 043
[arXiv:1009.1877 [hep-th]].



\bibitem{Saracco:2012wc}
F.~Saracco and A.~Tomasiello,
``Localized O6-plane solutions with Romans mass'',
JHEP \textbf{07} (2012), 077
[arXiv:1201.5378 [hep-th]].


\bibitem{Font:2019uva}
A.~Font, A.~Herr\'aez and L.~E.~Ib\'a\~nez,
``On scale separation in type II AdS flux vacua'',
JHEP \textbf{03} (2020), 013
[arXiv:1912.03317 [hep-th]]. 


\bibitem{Junghans:2020acz}
D.~Junghans,
``O-Plane Backreaction and Scale Separation in Type IIA Flux Vacua'',
Fortsch. Phys. \textbf{68} (2020) no.6, 2000040
[arXiv:2003.06274 [hep-th]].


\bibitem{Buratti:2020kda}
G.~Buratti, J.~Calderon, A.~Mininno and A.~M.~Uranga,
``Discrete Symmetries, Weak Coupling Conjecture and Scale Separation in AdS Vacua'',
JHEP \textbf{06} (2020), 083
[arXiv:2003.09740 [hep-th]].


\bibitem{Marchesano:2020qvg}
F.~Marchesano, E.~Palti, J.~Quirant and A.~Tomasiello,
``On supersymmetric AdS$_{4}$ orientifold vacua'',
JHEP \textbf{08} (2020), 087
[arXiv:2003.13578 [hep-th]].


\bibitem{Baines:2020dmu}
S.~Baines and T.~Van Riet,
``Smearing orientifolds in flux compactifications can be OK'',
Class. Quant. Grav. \textbf{37} (2020) no.19, 195015
[arXiv:2005.09501 [hep-th]].


\bibitem{DeLuca:2021mcj}
G.~B.~De Luca and A.~Tomasiello,
``Leaps and bounds towards scale separation'',
JHEP \textbf{12} (2021), 086
[arXiv:2104.12773 [hep-th]].


\bibitem{Cribiori:2021djm}
N.~Cribiori, D.~Junghans, V.~Van Hemelryck, T.~Van Riet and T.~Wrase,
``Scale-separated AdS4 vacua of IIA orientifolds and M-theory'',
Phys. Rev. D \textbf{104} (2021) no.12, 126014
[arXiv:2107.00019 [hep-th]].


\bibitem{Andriot:2022yyj}
D.~Andriot, L.~Horer and P.~Marconnet,
``Exploring the landscape of (anti-) de Sitter and Minkowski solutions: group manifolds, stability and scale separation'',
JHEP \textbf{08} (2022), 109
[erratum: JHEP \textbf{09} (2022), 184]
[arXiv:2204.05327 [hep-th]].


\bibitem{Andriot:2022brg}
D.~Andriot, L.~Horer and G.~Tringas,
``Negative scalar potentials and the swampland: an Anti-Trans-Planckian Censorship Conjecture'',
JHEP \textbf{04} (2023), 139
[arXiv:2212.04517 [hep-th]].



\bibitem{Shiu:2022oti}
G.~Shiu, F.~Tonioni, V.~Van Hemelryck and T.~Van Riet,
``AdS scale separation and the distance conjecture'',
JHEP \textbf{05} (2023), 077
[arXiv:2212.06169 [hep-th]].





\bibitem{Basile:2023rvm}
I.~Basile and C.~Montella,
``Domain walls and distances in discrete landscapes'' [arXiv:2309.04519 [hep-th]].




\bibitem{Andriot:2023fss}
D.~Andriot and G.~Tringas,
``Extensions of a scale separated AdS$_4$ solution and their mass spectrum'' [arXiv:2310.06115 [hep-th]].


\bibitem{Junghans:2023yue}
D.~Junghans,
``A note on O6 intersections in AdS flux vacua'' [arXiv:2310.17695 [hep-th]].


\bibitem{Conlon:2021cjk}
J.~P.~Conlon, S.~Ning and F.~Revello,
``Exploring the holographic Swampland'',
JHEP \textbf{04} (2022), 117
[arXiv:2110.06245 [hep-th]].

\bibitem{Apers:2022zjx}
F.~Apers, M.~Montero, T.~Van Riet and T.~Wrase,
``Comments on classical AdS flux vacua with scale separation'',
JHEP \textbf{05} (2022), 167
[arXiv:2202.00682 [hep-th]].


\bibitem{Apers:2022tfm}
F.~Apers, J.~P.~Conlon, S.~Ning and F.~Revello,
``Integer conformal dimensions for type IIa flux vacua'',
Phys. Rev. D \textbf{105} (2022) no.10, 106029
[arXiv:2202.09330 [hep-th]].


\bibitem{Quirant:2022fpn}
J.~Quirant,
``Noninteger conformal dimensions for type IIA flux vacua'',
Phys. Rev. D \textbf{106} (2022) no.6, 066017
[arXiv:2204.00014 [hep-th]].


\bibitem{Plauschinn:2022ztd}
E.~Plauschinn,
``Mass spectrum of type IIB flux compactifications \textemdash{} comments on AdS vacua and conformal dimensions'',
JHEP \textbf{02} (2023), 257
[arXiv:2210.04528 [hep-th]].


\bibitem{Apers:2022vfp}
F.~Apers,
``Aspects of AdS flux vacua with integer conformal dimensions'',
JHEP \textbf{05} (2023), 040
[arXiv:2211.04187 [hep-th]].





\bibitem{Carrasco:2023hta}
R.~Carrasco, T.~Coudarchet, F.~Marchesano and D.~Prieto,
``New families of scale separated vacua'' [arXiv:2309.00043 [hep-th]].


\bibitem{Tringas:2023vzn}
G.~Tringas,
``Anisotropic scale-separated AdS$_4$ flux vacua'' [arXiv:2309.16542 [hep-th]].


\bibitem{Petrini:2013ika}
M.~Petrini, G.~Solard and T.~Van Riet,
``AdS vacua with scale separation from IIB supergravity'',
JHEP \textbf{11} (2013), 010
[arXiv:1308.1265 [hep-th]].


\bibitem{Emelin:2021gzx}
M.~Emelin, F.~Farakos and G.~Tringas,
``Three-dimensional flux vacua from IIB on co-calibrated G2 orientifolds'',
Eur. Phys. J. C \textbf{81} (2021) no.5, 456
[arXiv:2103.03282 [hep-th]].



\bibitem{Danielsson:2018ztv}
U.~H.~Danielsson and T.~Van Riet,
``What if string theory has no de Sitter vacua?'',
Int. J. Mod. Phys. D \textbf{27} (2018) no.12, 1830007
[arXiv:1804.01120 [hep-th]].

\bibitem{Gautason:2018gln}
F.~F.~Gautason, V.~Van Hemelryck and T.~Van Riet,
``The Tension between 10D Supergravity and dS Uplifts'',
Fortsch. Phys. \textbf{67} (2019) no.1-2, 1800091
[arXiv:1810.08518 [hep-th]].

\bibitem{Hamada:2018qef}
Y.~Hamada, A.~Hebecker, G.~Shiu and P.~Soler,
``On brane gaugino condensates in 10d'',
JHEP \textbf{04} (2019), 008
[arXiv:1812.06097 [hep-th]].


\bibitem{Gao:2020xqh}
X.~Gao, A.~Hebecker and D.~Junghans,
``Control issues of KKLT'',
Fortsch. Phys. \textbf{68} (2020), 2000089
[arXiv:2009.03914 [hep-th]].


\bibitem{Emelin:2020buq}
M.~Emelin,
``Effective Theories as Truncated Trans-Series and Scale Separated Compactifications'',
JHEP \textbf{11} (2020), 144
[arXiv:2005.11421 [hep-th]].




\bibitem{Green:2007zzb}
M.~B.~Green, H.~Ooguri and J.~H.~Schwarz,
``Nondecoupling of Maximal Supergravity from the Superstring'',
Phys. Rev. Lett. \textbf{99} (2007), 041601
[arXiv:0704.0777 [hep-th]].

\bibitem{Cribiori:2022trc}
N.~Cribiori and G.~Dall'Agata,
``Weak gravity versus scale separation'',
JHEP \textbf{06} (2022), 006
[arXiv:2203.05559 [hep-th]].



\bibitem{Cribiori:2023ihv}
N.~Cribiori and C.~Montella,
``Quantum gravity constraints on scale separation and de Sitter in five dimensions'' [arXiv:2303.04162 [hep-th]].




\bibitem{Cribiori:2023gcy}
N.~Cribiori and F.~Farakos,
``Supergravity EFTs and swampland constraints'',
PoS \textbf{CORFU2022} (2023), 167
[arXiv:2304.12806 [hep-th]].




\bibitem{Apruzzi:2021nle}
F.~Apruzzi, G.~Bruno De Luca, G.~Lo Monaco and C.~F.~Uhlemann,
``Non-supersymmetric AdS$_{6}$ and the swampland'',
JHEP \textbf{12} (2021), 187
[arXiv:2110.03003 [hep-th]].



\bibitem{Apruzzi:2019ecr}
F.~Apruzzi, G.~Bruno De Luca, A.~Gnecchi, G.~Lo Monaco and A.~Tomasiello,
``On AdS$_{7}$ stability'',
JHEP \textbf{07} (2020), 033
[arXiv:1912.13491 [hep-th]].



\bibitem{Farakos:2020phe}
F.~Farakos, G.~Tringas and T.~Van Riet,
``No-scale and scale-separated flux vacua from IIA on G2 orientifolds'',
Eur. Phys. J. C \textbf{80} (2020) no.7, 659
[arXiv:2005.05246 [hep-th]].


\bibitem{Emelin:2022cac}
M.~Emelin, F.~Farakos and G.~Tringas,
``O6-plane backreaction on scale-separated Type IIA AdS$_{3}$ vacua'',
JHEP \textbf{07} (2022), 133
[arXiv:2202.13431 [hep-th]].


\bibitem{VanHemelryck:2022ynr}
V.~Van Hemelryck,
``Scale-Separated AdS3 Vacua from G2-Orientifolds Using Bispinors'',
Fortsch. Phys. \textbf{70} (2022) no.12, 2200128
[arXiv:2207.14311 [hep-th]].


\bibitem{Farakos:2023nms}
F.~Farakos, M.~Morittu and G.~Tringas,
``On/off scale separation'',
JHEP \textbf{10} (2023), 067
[arXiv:2304.14372 [hep-th]].


\bibitem{Dibitetto:2018ftj}
G.~Dibitetto, G.~Lo Monaco, A.~Passias, N.~Petri and A.~Tomasiello,
``AdS$_3$ Solutions with Exceptional Supersymmetry'',
Fortsch. Phys. \textbf{66} (2018) no.10, 1800060
[arXiv:1807.06602 [hep-th]].


\bibitem{Passias:2020ubv}
A.~Passias and D.~Prins,
``On supersymmetric AdS$_{3}$ solutions of Type II'',
JHEP \textbf{08} (2021), 168
[arXiv:2011.00008 [hep-th]].


\bibitem{Macpherson:2021lbr}
N.~T.~Macpherson and A.~Tomasiello,
``$ \mathcal{N} $ = (1, 1) supersymmetric AdS$_{3}$ in 10 dimensions'',
JHEP \textbf{03} (2022), 112
[arXiv:2110.01627 [hep-th]].



\bibitem{VanRiet:2023pnx}
T.~Van Riet and G.~Zoccarato,
``Beginners lectures on flux compactifications and related Swampland topics'' [arXiv:2305.01722 [hep-th]].



\bibitem{Coudarchet:2023mfs}
T.~Coudarchet,
``Hiding the extra dimensions: A review on scale separation in string theory'',
[arXiv:2311.12105 [hep-th]].



\bibitem{Shiu:2023bay}
G.~Shiu, F.~Tonioni, V.~Van Hemelryck and T.~Van Riet,
``Connecting flux vacua through scalar field excursions''
[arXiv:2311.10828 [hep-th]].





\bibitem{Li:2023gtt}
Y.~Li, E.~Palti and N.~Petri,
``Towards AdS distances in string theory'',
JHEP \textbf{08} (2023), 210
[arXiv:2306.02026 [hep-th]].


\bibitem{Klaewer:2018yxi}
D.~Klaewer, D.~L\"ust and E.~Palti,
``A Spin-2 Conjecture on the Swampland'',
Fortsch. Phys. \textbf{67} (2019) no.1-2, 1800102
[arXiv:1811.07908 [hep-th]].



\bibitem{Higuchi:1986py}
A.~Higuchi,
``Forbidden Mass Range for Spin-2 Field Theory in De Sitter Space-time'',
Nucl. Phys. B \textbf{282} (1987), 397-436.





\end{thebibliography}
\end{document}